\documentclass[lettersize,journal]{IEEEtran}
\ifCLASSINFOpdf
\else
\usepackage{graphicx}

\graphicspath{{../eps/}}

\DeclareGraphicsExtensions{.eps}
\fi

\usepackage{amsmath}
\usepackage{amsfonts,amssymb}
\usepackage{amssymb}
\usepackage{wrapfig}
\usepackage{psfrag}
\usepackage{epstopdf}
\usepackage{graphicx}
\usepackage{float} 
\usepackage{subcaption}
\usepackage{threeparttable}
\usepackage{cases}
\usepackage{subeqnarray}
\usepackage{color}
\usepackage{underscore}
\usepackage{verbatim}
\usepackage{bm}
\usepackage{stfloats}
\usepackage{ragged2e} 
\usepackage{diagbox}
\usepackage{multirow}     
\usepackage{longtable}
\usepackage{url}
\usepackage[normalem]{ulem}
\usepackage{tabularray}

\usepackage{algorithm}
\usepackage{algorithmicx}
\usepackage{algpseudocode}
\usepackage{makecell}
\usepackage{authblk}
\usepackage[style=nature, backend=biber]{biblatex}
\begin{document}
\title{Paradigm Shift from Statistical Channel

Modeling to Digital Twin Prediction:

An Environment-Generalizable ChannelLM

for 6G AI-enabled Air Interface}


\author[1]{Yichen Cai}
\author[1]{Yuelong Qiu}
\author[1]{Jianhua Zhang}
\author[1]{Li Yu}
\author[1]{Yuxiang Zhang}
\author[2]{Zhen Zhang}
\author[3]{Guangyi Liu}

\affil[1]{State Key Lab of Networking and Switching Technology, Beijing University of Posts and Telecommunications, Beijing 100876, China}
\affil[2]{Inner Mongolia Key Laboratory of Intelligent Communication and Sensing and Signal Processing, Inner Mongolia University, Hohhot 010021, China}
\affil[3]{Future Research Laboratory, China Mobile Research Institute, Beijing 100053, China}

\affil[ ]{\small
caiyichen@bupt.edu.cn,
yl\_qiu@bupt.edu.cn,
jhzhang@bupt.edu.cn\textsuperscript{*},
li.yu@bupt.edu.cn\textsuperscript{*}\\
zhangyx@bupt.edu.cn,
zhenzhang@imu.edu.cn,
liuguangyi@chinamobile.com\textsuperscript{*}
}
\affil[ ]{\small
Corresponding authors: Jianhua Zhang, Li Yu, and Guangyi Liu
}

\maketitle

\noindent\textbf{Abstract}

As 6G advances, ubiquitous connectivity and higher capacity requirements of the air interface pose substantial challenges for accurate and real-time wireless channel acquisition in diverse environments. Conventional statistical channel modeling relies on offline measurement data from limited environments, struggling to support online applications facing diverse environments. To this end, the digital twin channel (DTC) has emerged as a novel paradigm that constructs a digital replica of the physical environment through high-fidelity sensing and predicts corresponding channel in real time utilizing artificial intelligence (AI) models. As the engine of DTC, existing AI models struggle to simultaneously achieve strong environmental generalization in real-world and end-to-end channel prediction for real time tasks. Therefore, this paper proposes a channel large model (ChannelLM)-driven DTC architecture comprising three modules: low-complexity and high-accuracy environment reconstruction based on dynamic object detection and multimodal alignment of image and point cloud data, physically interpretable environment feature extraction, and a ChannelLM core to mapping these features into generalized environment representations for multi-task channel prediction. Simulation results demonstrate that, in unseen test environments, compared with small-scale AI models, ChannelLM reduces prediction errors by 4.23 dB in channel state information prediction while achieving an end-to-end inference latency of 70 milliseconds in the real world. 

\IEEEpeerreviewmaketitle
\vspace{4mm}
\noindent\textbf{Introduction}

Toward the 2030 commercial vision, the international telecommunication union radiocommunication sector (ITU-R) has defined typical scenarios for the sixth generation (6G), such as ubiquitous connectivity and integrated artificial intelligence (AI) and communications, while imposing stringent air interface requirements including terabit-per-second (Tbps) peak data rates \cite{itu2160, WEIT, 6G1, zpAI, AIAI}. To achieve these key performance indicator across diverse 6G scenarios such as the internet of vehicles (IoV) and low-altitude communications, the accurate and adaptable channel models are critical \cite{zjh6G, ZZ_AI, zz_drl,ztAI}. However, conventional statistical channel modeling essentially relies on collecting offline channel impulse response (CIR) data within limited predefined typical scenarios (e.g., urban macrocells, indoor hotspots) to fit statistical distributions of random parameters such as delay and angle~\cite{38901}. The fundamental limitation of statistical channel modeling lies in its insufficient environmental generalization capability: when confronted with transitions between different scenarios of dynamic environmental variations. Consequently, there is an urgent
need to explore new paradigms for accurate and online
channel prediction with high generalization capability, to support the intelligent evolution of air interface technologies~\cite{Cluster-nuclei}.

As wireless channel characteristics are intrinsically environment-dependent, a profound exploration of the physical environment has emerged as a pivotal breakthrough for advancing new channel prediction paradigms. Concurrently, the rapid evolution of integrated sensing and communication (ISAC) technologies in 6G systems, coupled with the ubiquitous deployment of multi-modal sensing infrastructures, has enabled the rapid sensing and reconstruction of complex environments. Prior studies have achieved high-fidelity environment reconstruction through the joint scheduling of multi-modal sensors, including red-green-blue (RGB) cameras, depth cameras, and light detection and ranging (LiDAR), thereby fusing visual, depth, and point cloud information to obtain multi-dimensional complementary environmental representations \cite{ESandReC1, ESandReC2, ESandReC3}. This process serves as a prerequisite for constructing environment-generalizable channel prediction paradigms~\cite{radar-com}.

Benefiting from efficient environment sensing capabilities in 6G systems and rapid advances in AI technologies, the digital twin channel (DTC) has emerged in recent years as a novel paradigm for channel prediction~\cite{DTC3,DTC4,DTC5,ylbupt,4steps,slzmwef}. In our prior work~\cite{REKP, DTC2}, we first define DTC as a channel acquisition approach capable of real-time mapping of channel fading states and their dynamic evolution. Through continuous interaction with system requirements, it enables communication systems to proactively adapt to environmental changes. Its implementation pipeline comprises: multi-modal environment sensing data collection, processing, feature extraction, channel prediction, and subsequent communication optimization. Compared with conventional statistical channel modeling, the DTC paradigm breaks through the constraints of typical scenarios such as urban macrocells and indoor hotspots, accommodating more complex and diverse practical environments. Moreover, its online update mechanism based on real-time environment sensing overcomes the limitations of traditional offline statistical fitting.

Despite the considerable attention garnered by DTC-based wireless channel prediction in both industry and academia, numerous challenges persist in coping with the complex environmental variations of 6G. At the industrial implementation level, advanced platforms, such as NVIDIA Aerial OmniGraph Digital Twin (AODT), still rely primarily on prior high-precision three-dimensional (3D) environment maps and accelerated ray tracing~\cite{AODT1, AODT2}, rendering them inadequate for the precise and efficient prediction of wireless channels amidst real-time environmental changes in 6G systems.

At the academic research level, existing channel prediction methods based on single/multi-modal environment sening information and conventional AI small models, such as convolutional neural networks (CNNs) and U-shaped networks (UNets), are typically confined to specific environments or single prediction tasks, exhibiting limited environmental generalization capability~\cite{SM1, MM1, MM2, MM3, MM5, MM7,MM9}. Although AI large models built upon large language models (LLMs) have introduced new opportunities for channel prediction, current mainstream research remains focused on direct processing of channel data, lacking explicit modeling of wireless environment that fundamentally governs channel characteristics~\cite{LM00,LM0, LM1, LM2, LM3, LM4, LM5, LM6}. The few works incorporating wireless environment have merely operated at the raw data level of images and point clouds~\cite{LM7}, failing to perform interpretable feature extraction tailored to radio propagation properties. Consequently, these models struggle to effectively suppress redundant information and environmental noise, severely constraining their performance potential.

Based on the above analysis, conventional statistical channel modeling and existing channel prediction schemes struggle to reconcile environmental generalization with computational efficiency. Specifically, they fail to simultaneously achieve: (i) robust generalization across diverse 6G scenarios, and (ii) real-time prediction of wireless channels.

To address these challenges, this paper proposes a novel DTC implementation based on channel large model (ChannelLM). The architecture comprises three core modules: multi-modal sensing based environment reconstruction module, radio propagation guided feature extraction module, and ChannelLM driven digital twin prediction module. Within this unified architecture, we achieve, for the first time, the joint prediction of large scale fading such as path loss and small scale information such as channel state information (CSI) in dynamic scenarios.

Specifically, to enhance real-time deployment capability, the multi-modal sensing-based environment reconstruction module employs a hierarchical reconstruction strategy that decomposes wireless environment into static and dynamic components, shifting the computational focus from full-scene reconstruction to the real-time updating of sparse dynamic objects. By fusing red-green-blue-depth (RGB-D) images and LiDAR point cloud data, this scheme detects and reconstructs dynamic components in real time, which are subsequently integrated with pre-built static component to generate complete wireless environments. To improve the generalization ability across diverse scenarios, the radio propagation-guided environment feature extraction module extracts key physical features, including transmitter and receiver positions, scatterer distribution, and a penetration ratio proposed for the first time in this work. This effectively suppresses interference from irrelevant environmental variations on channel estimation. Furthermore, in the ChannelLM driven digital twin prediction module, a multi-scale patch design is adopted to capture both the global structure and local variations of the environment, while a pretrained large model, such as GPT-2, is used as the backbone to learn representations from the above interpretable features. This enables the model to extract more stable environment–channel correlation patterns and improves its adaptability to unseen environments. Then, collaborative prediction branches are designed to combine generalized environment representations with lightweight pilot inputs, enabling the prediction of PL maps and CSI matrices at specific locations within a unified architecture. 

Simulation results show that the proposed ChannelLM achieves excellent generalization performance in previously unseen test environments for the collaborative prediction of large-scale PL and small-scale CSI matrices. Furthermore, in dynamic scenarios caused by pedestrian movement, the system level end-to-end inference latency encompassing sensing, feature extraction, and digital twin prediction can be controlled at approximately 70 milliseconds (ms), satisfying real-time prediction requirements. The above simulation results further demonstrate that the proposed architecture can support the intelligent evolution of the AI-enabled air interface, in which the ability to adapt to environmental dynamics fundamentally relies on real-time and reliable channel acquisition. The above simulation results further demonstrate that the proposed architecture enables real-time, adaptive channel prediction in dynamic 6G environments, paving the way for a paradigm shift from conventional statistical channel modeling to DTC and thereby supporting the intelligent evolution of the AI-enabled air interface.

\begin{figure*}[t]
\centerline{\includegraphics[width=1.0\textwidth]{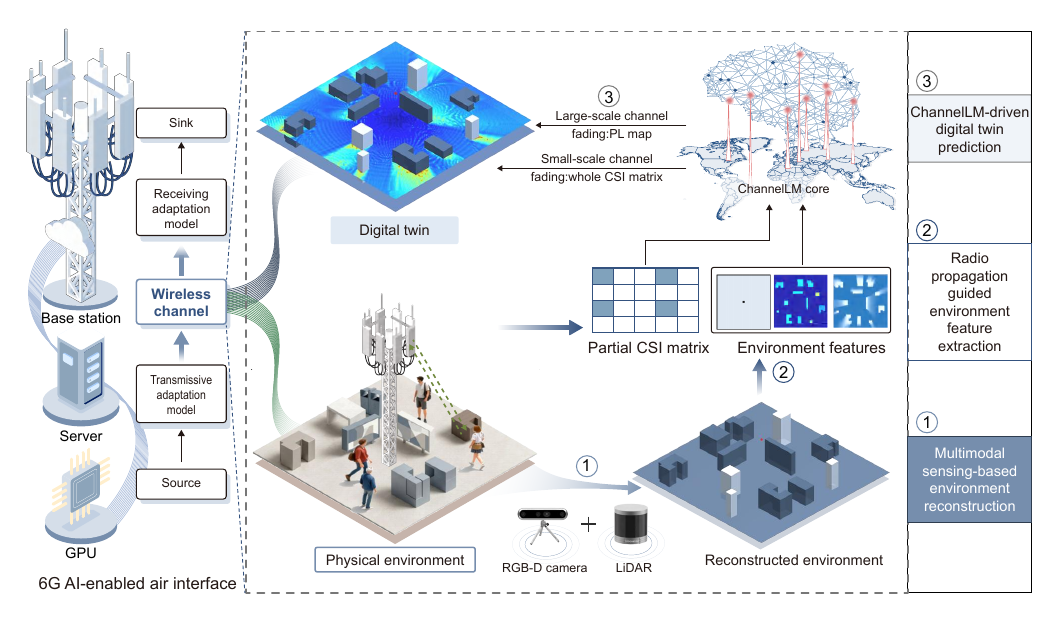}}
\caption{A ChannelLM-based implementation of DTC paradigm for the 6G AI-enabled air interface}
\label{FigChannelLMFramework}
\end{figure*}

\vspace{4mm}
\noindent\textbf{Results}

\noindent\textbf{System Model}

Consider a downlink wireless communication system employing multiple-input single-output orthogonal frequency-division multiplexing (MISO-OFDM) with $N_K$ subcarriers, where the base station (BS) is equipped with a uniform linear array (ULA) consisting of $N_t$ antenna elements, transmitting OFDM-modulated signals to a single-antenna user terminal (UT). For any subcarrier $k = 1, 2,..., N_K$, the corresponding downlink CSI vector can be modeled as:
\begin{equation}
    \mathbf{h}\left(k\right)=\sum_{l=1}^{L}\alpha_{l}e^{-j2\pi k \Delta f\tau_l}\mathbf{a}_k\left(\phi_l\right),
    \label{CSIVector}
\end{equation}
where $\mathbf{h}\left(k\right)\in\mathbb{C}^{N_t}$ is the downlink CSI vector of subcarrier $k$. The parameters $\alpha_{l}$, $\tau_{l}$ and $\phi_{l}$ are the complex gain (encompasses both amplitude fading and phase shift), delay and angle of departure of the $l$-th path, respectively. Moreover, $\Delta f$ is the subcarrier spacing, and $\mathbf{a}_k\left(\phi_l\right)$ is the steering vector of ULA for subcarrier $k$, which can be expressed as:
\begin{equation}
    \mathbf{a}_k\left(\phi_l\right)=\left[1,e^{-j\frac{2\pi d }{\lambda_k}\sin\phi_l},...,e^{-j\frac{2\pi d }{\lambda_k}\left(N_t-1\right)\sin\phi_l}\right]^T,
    \label{SteeringVector}
\end{equation}
where $d$ is the antenna spacing in the ULA, and $\lambda_k$ is the wavelength of subcarrier $k$.

The complete space-frequency CSI matrix $\mathbf{H}\in\mathbb{C}^{N_t\times K}$ is obtained by concatenating the CSI vector of all $K$ subcarriers, which can be expressed as:
\begin{equation}
    \mathbf{H}=\left[\mathbf{h}\left(1\right), \mathbf{h}\left(2\right),..., \mathbf{h}\left(K\right)\right].
    \label{CSIMatrix}
\end{equation}
Based on the space-frequency channel model for MISO-OFDM system given in equation \eqref{CSIMatrix}, the large-scale PL in dB can be calculated as follows:
\begin{equation}
    PL_{dB}=-10\log_{10}\left(\frac{1}{N_tK}\left|\left|\mathbf{H}\right|\right|_F^2\right),
    \label{PLModel}
\end{equation}
where $\left|\left|\cdot\right|\right|_F$ is the Frobenius norm of a matrix.

\vspace{4mm}
\noindent\textbf{Problem Formulation}

Based on the above CSI and PL models, this paper focuses on digital twin prediction in complex, dynamic scenarios, including PL map and CSI matrix predictions. For PL map prediction, two conventional approaches are commonly employed: direct calculation from the ratio of transmitted to received signal power, and derivation via equation \eqref{PLModel} after obtaining the whole CSI matrix $\mathbf{H}$. However, both strategies are limited by two key issues. Firstly, in MISO-OFDM systems, they typically require a large number of pilot resources overhead. Secondly, the PL map prediction is confined to a specific UT location $\mathbf{q}_{UT}$, as only the direct link between the BS and UT at that particular position is considered.

\vspace{4mm}
\noindent\textbf{Environment Reconstruction and Feature Extraction}\label{sec_env}

Inspired by the concepts from DTC and image processing, it is promising to use BS location and wireless environment information as input to an AI model to directly reconstruct a complete PL map covering all locations within the target communication area. The corresponding expression is as follows:
\begin{equation}
    \widehat{\mathbf{P}}=g_{PL}\left(\mathbf{q}_{BS},\mathcal{E}\right),
    \label{PLPreProblem}
\end{equation}
where $\widehat{\mathbf{P}}$, referred to as the PL map, represents the PL values at discrete locations across the target area, whose dimensions are determined by the physical size of the area and the resolution of the discretization grid. $\mathbf{q}_{BS}$ denotes the location of the BS, while $\mathcal{E}$ encompasses the set of relevant environment characteristics, including geometry distribution, height, etc. The function $g_{PL}(\cdot)$ corresponds to an AI model designed for predicting the PL map.

For space-frequency domain CSI matrix prediction, conventional pilot-based approaches incur prohibitive overhead, as the required number of orthogonal pilots scales with the number of antennas and subcarriers.  Wireless environmental information theory (WEIT) indicates that a deeper understanding of wireless environment reduces the information entropy of channels in corresponding scenarios, thereby improving prediction accuracy \cite{WEIT}. Furthermore, our prior work \cite{LM6} has shown that incorporating environment features can effectively reduce the pilot overhead required for CSI prediction, thereby enhancing communication efficiency. Thus, for any UT location $\mathbf{q}_{UT}$ within the target communication area, CSI matrix prediction combining environment features with lightweight pilot observations can be formulated as:
\begin{equation}                            
    \widehat{\mathbf{H}}\left(\mathbf{q}_{UT}\right)=g_{CSI}\left(\mathbf{q}_{BS},\mathbf{q}_{UT},\mathcal{E},\mathbf{H}_{0}\left(\mathbf{q}_{UT}\right)\right),
    \label{CSIPreProblem}
\end{equation}
where $\widehat{\mathbf{H}}\left(\mathbf{q}_{UT}\right)\in\mathbb{C}^{N_t\times K}$ and $\mathbf{H}_0\left(\mathbf{q}_{UT}\right)\in\mathbb{C}^{N_t\times K}$ are the whole CSI matrix to be recovered and the real-time CSI matrix obtained via lightweight spatial-frequency pilot measurements, respectively. Denote $g_{CSI}(\cdot)$ as an AI model designed for CSI matrix reconstruction. Additionally, $\mathbf{H}_0\left(\mathbf{q}_{UT}\right)$ and the channel matrix $\mathbf{H}\left(\mathbf{q}_{UT}\right)$ in equation \eqref{CSIPreProblem} satisfy the following relationship:
\begin{equation}
    \mathbf{H}_0\left(\mathbf{q}_{UT}\right)=\mathbf{A}\odot\mathbf{H}\left(\mathbf{q}_{UT}\right),
    \label{CSIMatrixZero}
\end{equation}
where $\mathbf{A}\in\mathbb{R}^{N_t\times K}$ is a sampling matrix whose elements are equal to one only at the spatial–frequency grid points corresponding to pilot positions and zero elsewhere; $\odot$ denotes the Hadamard product. This measurement mechanism allows $\mathbf{H}_0\left(\mathbf{q}_{UT}\right)$ to capture both large-scale and small-scale fading characteristics at location $\mathbf{q}_{UT}$ in complex dynamic environments, thereby effectively supplementing environmental prior information and improving the recovery accuracy of the channel matrix $\widehat{\mathbf{H}}\left(\mathbf{q}_{UT}\right)$ to be estimated in equation \eqref{CSIPreProblem}.

This paper proposes a ChannelLM-driven DTC architecture for the 6G AI-enabled air interface, as illustrated in Fig. \ref{FigChannelLMFramework}. The architecture comprises three core modules: 1) multi-modal sensing-based environment reconstruction, 2) radio propagation-guided environment feature extraction, and 3) ChannelLM-driven digital twin prediction, which perform environment sensing, environment understanding, and environment utilization, respectively. Therefore, a real-time digital twin prediction framework is established, linking physical environment information to wireless channel fading states, including large-scale fading represented by PL and small-scale fading represented by CSI.

For multi-modal sensing-based environment reconstruction, this paper proposes a hierarchical reconstruction procedure to balance the dual requirements of reconstruction accuracy and real-time performance, by integrating offline modeling of quasi-static environments with online sensing and reconstruction of dynamic object. Specifically, the complete environment information is decoupled into quasi-static components over the time scale (like buildings, roads, and vegetation) and dynamic components (like vehicles and pedestrians). A high-precision static environment is first established through offline precision measurement and 3D modeling. During the online phase, multi-modal sensors such as RGB-D cameras and LiDAR are utilized for real-time dynamic information acquisition. The RGB-D camera provides high-resolution RGB images along with corresponding depth data, while the LiDAR enhances localization accuracy in challenging scenarios such as strong light, backlight, and medium-to-long distances through high-precision discrete point clouds. By fusing these sensing data, real-time detection and 3D position estimation of dynamic objects can be achieved, which are then embedded into the static environment in the form of cuboid bounding boxes. Through deep fusion, a complete 3D environment reconstruction is achieved, incorporating both high-precision static structures and real-time dynamic objects. This reconstruction provides reliable input for subsequent environment feature extraction and digital twin prediction.

\begin{figure}[t]
\centerline{\includegraphics[width=0.45\textwidth]{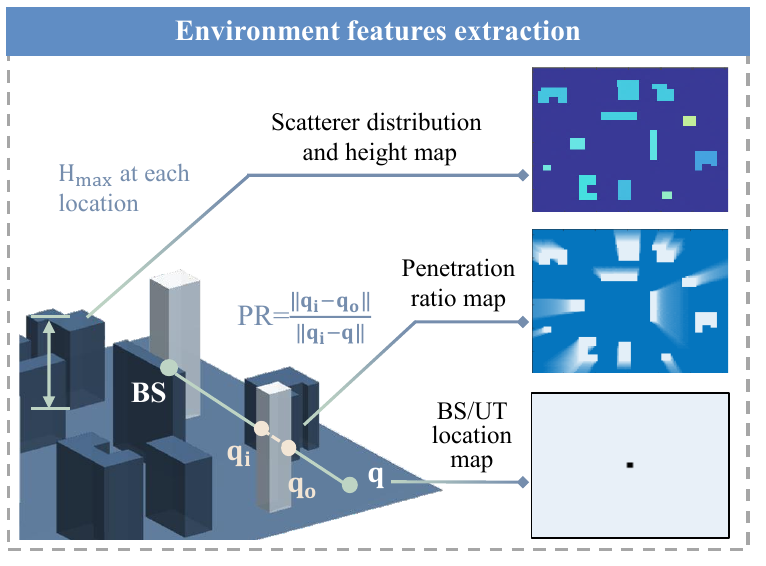}}
\caption{The details of the extracted environment feature.}
\label{FigEnvironmentFeature}
\end{figure}

For radio propagation-guided environment feature extraction, rather than relying solely on raw data, constructing informative and task-oriented feature representations is widely recognized as a key factor in the success of AI models, often making the learning problem more tractable and efficient \cite{featureEngineering}.
In line with the radio propagation-guided design principle, this paper performs systematic feature extraction from the reconstructed environment to identify key elements that are strongly correlated with channel propagation characteristics to enhance model efficacy. Specifically, the spatial locations of the transmitter and receiver, together with the geometric structure of the surrounding environment, play a important role in determining propagation behavior and the space-time-frequency response of the wireless channel. 
As illustrated in Fig.~\ref{FigChannelLMFramework}, this paper extract the following three types of environment features with clear physical interpretability: the locations of the BS and UT, the spatial distribution and height of scatterers, and the penetration ratio along the line connecting the BS and any target location, which quantifies the proportion of the line segment occupied by scatterers. 

The detailed definitions for each feature are elaborated below. To explicitly characterize the spatial relationship between the BS and UT in communication systems, this paper constructs a two-dimensional (2D) BS/UT location map for the selected communication scenario, employing a binarization approach to depict their relative location. Specifically, for any location $\mathbf{q} = [x, y]^T$ within the scenario, the binary state variables are defined as follows:
\begin{equation}
    B_{j}\left(\mathbf{q}\right)=\left\{\begin{array}{ll}
1, & \mathbf{q}=\mathbf{q}_{j} \\
0, & \text {otherwise }
\end{array}\right., j=BS,UT,
\label{BBSUTq}
\end{equation}
where $B_{BS}$ and $B_{UT}$ are the binary location maps of the BS and UT, respectively, illustrated in Fig. \ref{FigEnvironmentFeature}. Besides, $\mathbf{q}_{BS}$ and $\mathbf{q}_{UT}$ are the 2D location of the BS and UT, respectively. This explicit encoding enables the proposed ChannelLM to obtain the spatial positions of the transmitter and receiver in the most direct manner, and as a result, improves its performance of channel fading prediction.

Since multipath propagation mechanisms, such as reflection, scattering, and diffraction, are closely related to the spatial distribution and height attributes of scatterers in the physical environment, this paper constructs a scatterer distribution and height map for the selected communication scenario. Specifically, this map quantifies the vertical height of scatterers at each location in the 2D space relative to a horizontal ground plane. For any location $\mathbf{q} = [x, y]^T$ within the scenario, the height value $H\left(\mathbf{q}\right)$ is defined as follows:
\begin{equation}
    H\left(\mathbf{q}\right)=\left\{\begin{array}{ll}
h_{max}\left(\mathbf{q}\right), & \mathbf{q}\in\mathcal{S} \\
\quad 0, & \text {otherwise }
\end{array}\right.,
\label{Hq}
\end{equation}
where $H$ is the constructed scatterer distribution and height map illustrated in Fig. \ref{FigEnvironmentFeature}, $\mathcal{S}$ is the set of the 2D plane coordinates inside all scatterers, and $h_{max}\left(\mathbf{q}\right)$ is the maximum height of the scatterer at location $\mathbf{q}$. The scatterer distribution and height map provides a structured geometric representation of the propagation environment, which is helpful for predicting channel fading behaviors related to the presence and layout of scatterers.

\begin{figure*}[t]
\centerline{\includegraphics[width=0.8\textwidth]{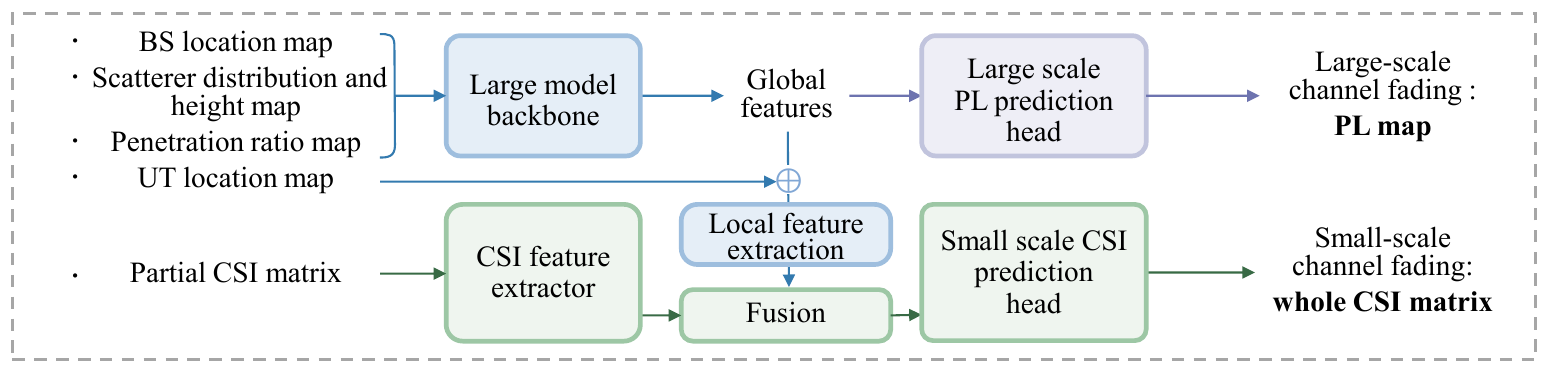}}
\caption{Architecture of the proposed channelLM core}
\label{FigLMarchitecture}
\end{figure*}

To quantify the degree of blockage between the BS and a target location, this paper introduces a penetration ratio map. In wireless communication systems, the signal propagation environments for BSs and UTs usually differ significantly. The BSs are typically deployed at elevated sites with low scatterer density in their vicinity. In contrast, UTs commonly operate at ground level in close proximity to dense scatterers. The resulting complexity in signal propagation, particularly the increased obstruction and attenuation experienced in non-line-of-sight (NLoS) paths, is mainly attributed to the dense scattering environment surrounding UTs. To quantitatively characterize such propagation blockage, the penetration ratio metric is introduced. For any potential UT location $\mathbf{q} = [x, y, z]^T$ within the selected communication scenario, penetration ratio can be defined as:
\begin{equation}
    PR\left(\mathbf{q}\right)=\left\{\begin{array}{ll}
\frac{\left|\left|\mathbf{q}_{i}-\mathbf{q}_{o}\right|\right|}{\left|\left|\mathbf{q}_{i}-\mathbf{q}\right|\right|}, & \text{LoS path not exists} \\
\qquad 0, & \text {otherwise }
\end{array}\right.,
\label{PRq}
\end{equation}
where $PR\left(\mathbf{q}\right)$ is the penetration ratio at location $\mathbf{q}$. $\mathbf{q}_{i}= [x_{i}, y_{i}, z_{i}]^T$ and $\mathbf{q}_{o}= [x_{o}, y_{o}, z_{o}]^T$ are the scatterer intersections nearest to the BS and to location $\mathbf{q}$, respectively, along the line connecting the BS and the location $\mathbf{q}$. Denote $\left|\left|\cdot\right|\right|$ as the 2-norm operation of a vector.

With identical spatial resolution and spatial consistency, the above three types of environment feature maps together constitute the multi-dimensional input to the large model, which serves to boost its performance in learning the dynamics of wireless channels.

\vspace{4mm}
\noindent\textbf{ChannelLM-Driven Digital Twin Prediction}\label{sec_arc}

For ChannelLM-driven digital twin prediction, this paper proposes a unified large model architecture to achieve multi-task collaborative prediction. As illustrated in Fig.~\ref{FigLMarchitecture}, the architecture jointly addresses two key tasks: predicting the PL map of the entire selected area and reconstructing the full CSI matrix at any given location. The core innovation of the proposed large model lies in the integrated design of a geometry-aware Transformer backbone and two task-specific prediction heads. Through a hierarchical feature extraction mechanism, the architecture globally models the large-scale influence of scene geometry on signal propagation, while simultaneously capturing localized scattering and fading details around the UT. This enables the unified model to jointly represent and accurately predict both large-scale and small-scale channel fading states. The prediction procedures for PL maps and CSI matrices are detailed below.

The PL prediction in communication systems with fixed antenna configurations and operating frequency bands is predominantly determined by the wireless propagation environment and transmission distance. Accordingly, this paper selects the BS location map, scatterer distribution and height map, and penetration ratio map obtained during the environment feature extraction stage as input features for channelLM. As illustrated in Fig. \ref{FigLMarchitecture}, the large model backbone is first employed to perform deep extraction and encoding of global environment features. Subsequently, a lightweight convolutional neural networks (CNN) within the PL prediction head is used to predict the PL map for the target communication area.

Compared with traditional image feature extraction methods based on CNN, the Transformer architecture can establish direct associations between any locations, making it more suitable for analyzing the global impact of environment geometry on PL. Traditional CNNs typically rely on adjusting kernel sizes and increasing network depth to gradually expand their receptive field. In contrast, the Transformer first divides the image into patches of a specific size and then establishes global dependencies among all patches through the self-attention mechanism. Under this mechanism, each patch can directly interact with any other patch in the image, thereby possessing a truly global receptive field.

The CSI prediction, unlike the PL prediction that exhibits spatial smoothness in characterizing large-scale fading, exhibits rapid spatial fluctuations due to multipath coherent superposition. Consequently, precise spatial characterization of the UT and extraction of local environment features in its vicinity are critical for enhancing CSI prediction accuracy. Building on the global environment features obtained in the PL map prediction stage, ChannelLM incorporates UT location maps during CSI prediction to capture localized environment information, which is subsequently fused with CSI estimated from sparse pilots for precise channel recovery. Specifically, as illustrated in Fig. \ref{FigLMarchitecture}, the CSI feature extractor first performs initial reconstruction from limited pilot observations using a CNN integrated with proximal gradient algorithms, thereby extracting CSI latent features. The local feature extraction module then utilizes UT position information to distill location-specific features from the global feature, enabling the model to perceive propagation characteristics at specific locations. Next, the feature fusion module integrates CSI latent features with local features, conditioning the final CSI generation on both environment geometry priors and actual channel observations to improve prediction accuracy and reliability. Finally, the CSI prediction head employs a lightweight two-layer CNN to efficiently map the fused features to the final CSI prediction. 


\vspace{4mm}
\noindent\textbf{Prediction Performance on Static Simulation Scenarios}

In the following experiments, the large-model backbone in ChannelLM is instantiated using GPT-2 small, which contains 124M parameters. For PL map prediction, this paper adopts the mean root mean square error (RMSE) as the evaluation metric. The RMSE between the predicted PL map $\widehat{\mathbf{P}}$ and the ground-truth PL map $\mathbf{P}$ is defined as:
\begin{equation}
    \text{RMSE}\left(\widehat{\mathbf{P}},\mathbf{P}\right)=\sqrt{\frac{1}{N_{PL}}\sum_{n=1}^{N_{PL}}\left(\widehat{P}_n-P_n\right)^2},
    \label{RMSE}
\end{equation}
where $N_{PL}$ is the total number of discrete locations in the non-scattering region of the selected scenario where PL needs to be predicted. $\widehat{P}_n$ and $P_n$ are the predicted and the ground-true PL value at the $n$-th location, respectively, both measured in dB. The mean RMSE are then computed by statistical processing of $\text{RMSE}\left(\widehat{\mathbf{P}},\mathbf{P}\right)$ across all given test samples. 

Furthermore, this paper adopts the normalized mean square error (NMSE) and mean squared generalized cosine similarity (SGCS) as the metric for evaluating the CSI matrices inference results, which can be defined as:
\begin{equation}
    \text{NMSE}=\frac{1}{N_{test}}\sum_{n=1}^{N_{test}}\frac{\left|\left|\widehat{\mathbf{H}}_n-\mathbf{H}_n\right|\right|_F^2}{\left|\left|\mathbf{H}_n\right|\right|_F^2},
    \label{eqNMSE}
\end{equation}
\begin{equation}
    \text{SGCS}=\frac{1}{N_{test}}\sum_{n=1}^{N_{test}}\frac{1}{N_{k}}\sum_{k=1}^{N_{k}}\frac{\left |\widehat{\tilde{\mathbf{h}}}_{n,k}^{\mathrm{H}}\tilde{\mathbf{h}}_{n,k}\right |^{2}}{\left\|\widehat{\tilde{\mathbf{h}}}_{n,k}\right\|_{2}^{2}\left\|\tilde{\mathbf{h}}_{n,k}\right\|_{2}^{2}}.
    \label{SGCS}
\end{equation}
where $N_{test}$ is the number of samples in the test set. $\widehat{\mathbf{H}}_n\in\mathbb{C}^{N_t\times K}$ and $\mathbf{H}_n\in\mathbb{C}^{N_t\times K}$ are the predicted and ground-truth CSI matrices, respecively. Meanwhile, $\widehat{\tilde{\mathbf{h}}}_{n,k}\in\mathbb{C}^{N_t}$ and $\tilde{\mathbf{h}}_{n,k}\in\mathbb{C}^{N_t}$ are the CSI vectors on the $k$-th subcarrier of $\widehat{\mathbf{H}}_n$ and $\mathbf{H}_n$, respectively. Notably, NMSE comprehensively evaluates prediction performance for both magnitude and phase of CSI matrices. In contrast, CSI magnitude terms cancel out in the numerator and denominator of the SGCS expression, rendering it a dedicated metric for assessing small-scale phase prediction accuracy.

In PL map prediction, ChannelLM is compared with PMNet~\cite{PMNet} and an ablated version of ChannelLM (denoted as W/o GPT-2) to evaluate its prediction performance. For PMNet, the input includes the BS location map, scatterer distribution, and height map, consistent with its original paper~\cite{PMNet}. The W/o GPT-2 variant removes the GPT-2-based architecture and retains only the three-layer CNN for PL map prediction. The dataset is splited into training, validation, and test sets at a 7:1:2 ratio. Specifically, the 80\% of the scenarios are allocated to the training and validation sets, where the validation subset is utilized for optimal model selection and mitigating overfitting. And the remaining 20\% of scenarios are held out as the test set to assess the model's generalization to unseen data. 
\begin{figure}[t]
\centerline{\includegraphics[width=0.5\textwidth]{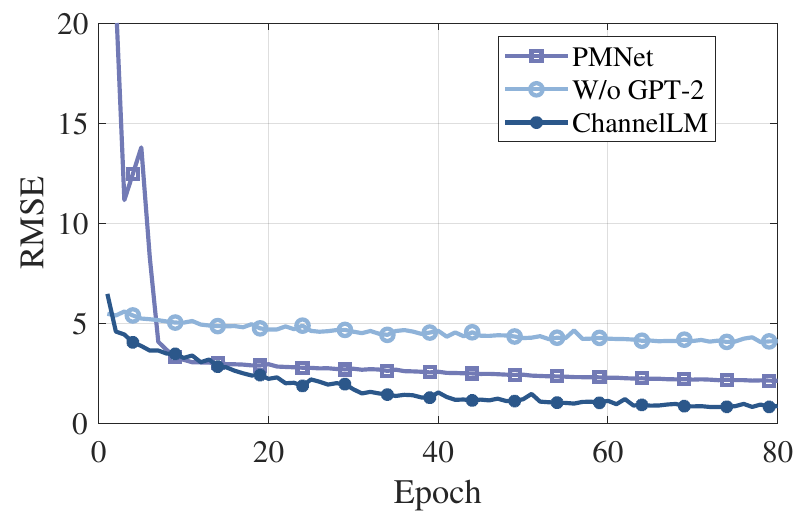}}
\caption{Mean RMSE and convergence trends of different models on the validation set.}
\label{FigPLRMSEEpoch}
\end{figure}

Fig. \ref{FigPLRMSEEpoch} illustrates the performance of these models on the validation set, which originates from the same pool of scenarios as the training set. It can be observed that for the $80$-th epoch, the mean RMSE values of the three models are 4.08, 2.12, and 0.80 dB, respectively, all achieving satisfactory fitting performance. Notably, the proposed ChannelLM exhibits faster convergence and lower error on the validation set, indicating its enhanced capability to extract environment features and thereby learn and characterize channel PL properties in specific scenarios. To further evaluate the generalization performance of the models in unseen scenarios, this paper tests all models on the test set that is independent of the training set. Table \ref{TablePLResult} presents the PL map prediction results for different models, with the mean and standard deviation of RMSE being obtained by statistical processing of all the $\text{RMSE}\left(\widehat{\mathbf{P}},\mathbf{P}\right)$ samples from test environments. 

ChannelLM achieves both the lowest mean RMSE and the smallest standard deviation, indicating that it not only attains a lower average error but also exhibits more consistent performance across different test regions. Compared with PMNet, ChannelLM reduces the mean RMSE by 1.43 dB; the reduction reaches 1.87 dB when compared with the W/o GPT-2, verifying the critical role of the sequence modeling module in capturing complex propagation mechanisms such as reflection, diffraction, and scattering. In terms of inference efficiency, the Transformer architecture’s attention and linear layers are highly parallelizable and can be efficiently accelerated via GPU matrix operations. Despite having significantly more parameters than PMNet, ChannelLM exhibits lower average inference latency, enhancing both prediction accuracy and stability without introducing notable time overhead, thus meeting real-time deployment requirements.


\begin{table}[t]
\centering
\caption{PL map prediction results for different models.}
\label{TablePLResult}
\renewcommand{\arraystretch}{1.5}
\begin{tabular}{|c|c|c|c|}
\hline
\rule{0pt}{4ex}\textbf{Model} & \textbf{\makecell[c]{Mean of\\ RMSE (dB)}} &\textbf{\makecell[c]{Model size\\(M)}} &\textbf{\makecell[c]{Average inference \\ latency (ms)}} \\
\hline
PMNet & 5.87 & 33.3 & 17.37  \\
\hline
W/o GPT-2& 6.31 & 0.43 & 0.5 \\
\hline
ChannelLM& 4.44 &140.6& 13.76 \\
\hline
\end{tabular}
\end{table}

\begin{figure*}[t]
    \centering
    \begin{subfigure}[t]{0.23\textwidth}
        \centering
        \includegraphics[width=\textwidth]{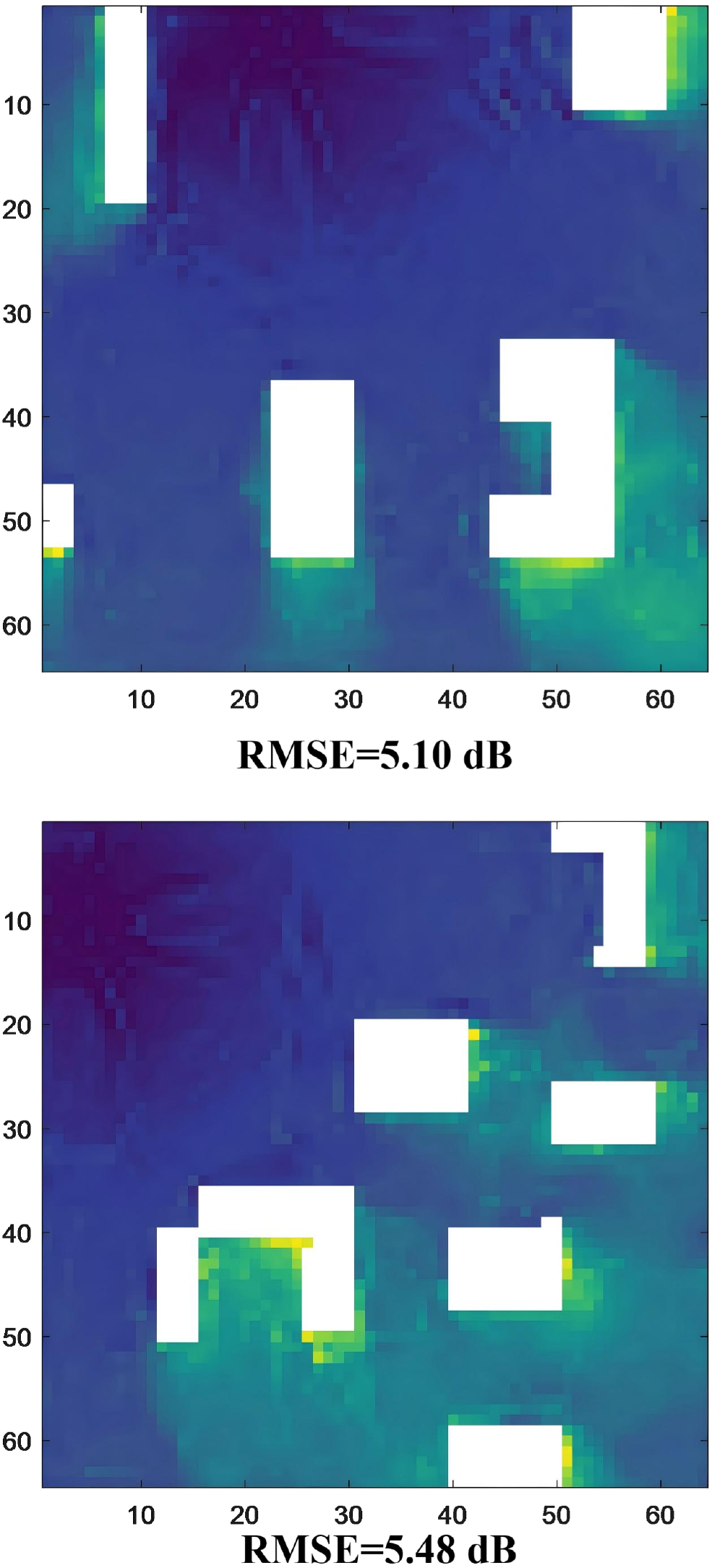}
        \caption{PMNet}
        \label{FigPLMapPreFigurea}
    \end{subfigure}
    \hfill
    \begin{subfigure}[t]{0.23\textwidth}
        \centering
        \includegraphics[width=\textwidth]{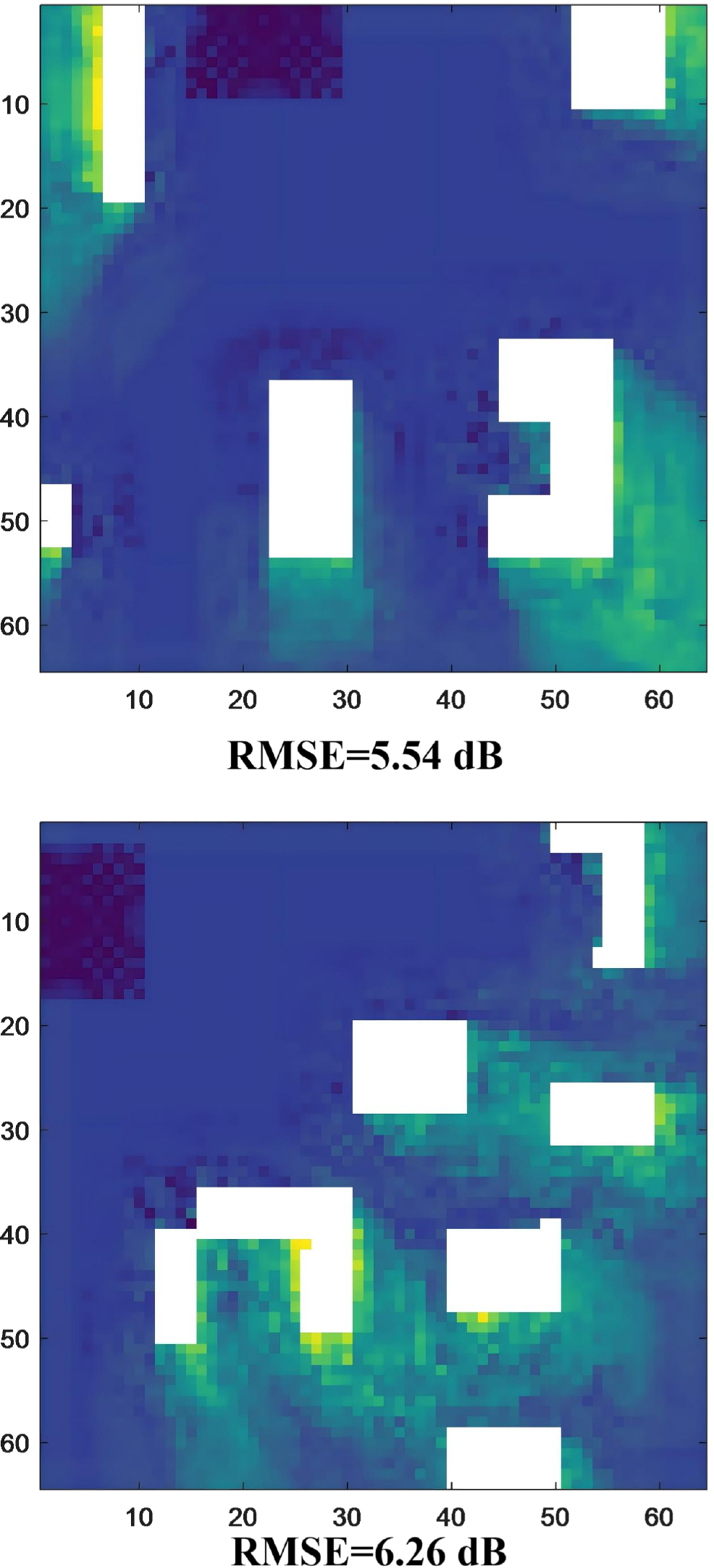}
        \caption{W/o GPT-2}
        \label{FigPLMapPreFigureb}
    \end{subfigure}
    \hfill
    \begin{subfigure}[t]{0.23\textwidth}
        \centering
        \includegraphics[width=\textwidth]{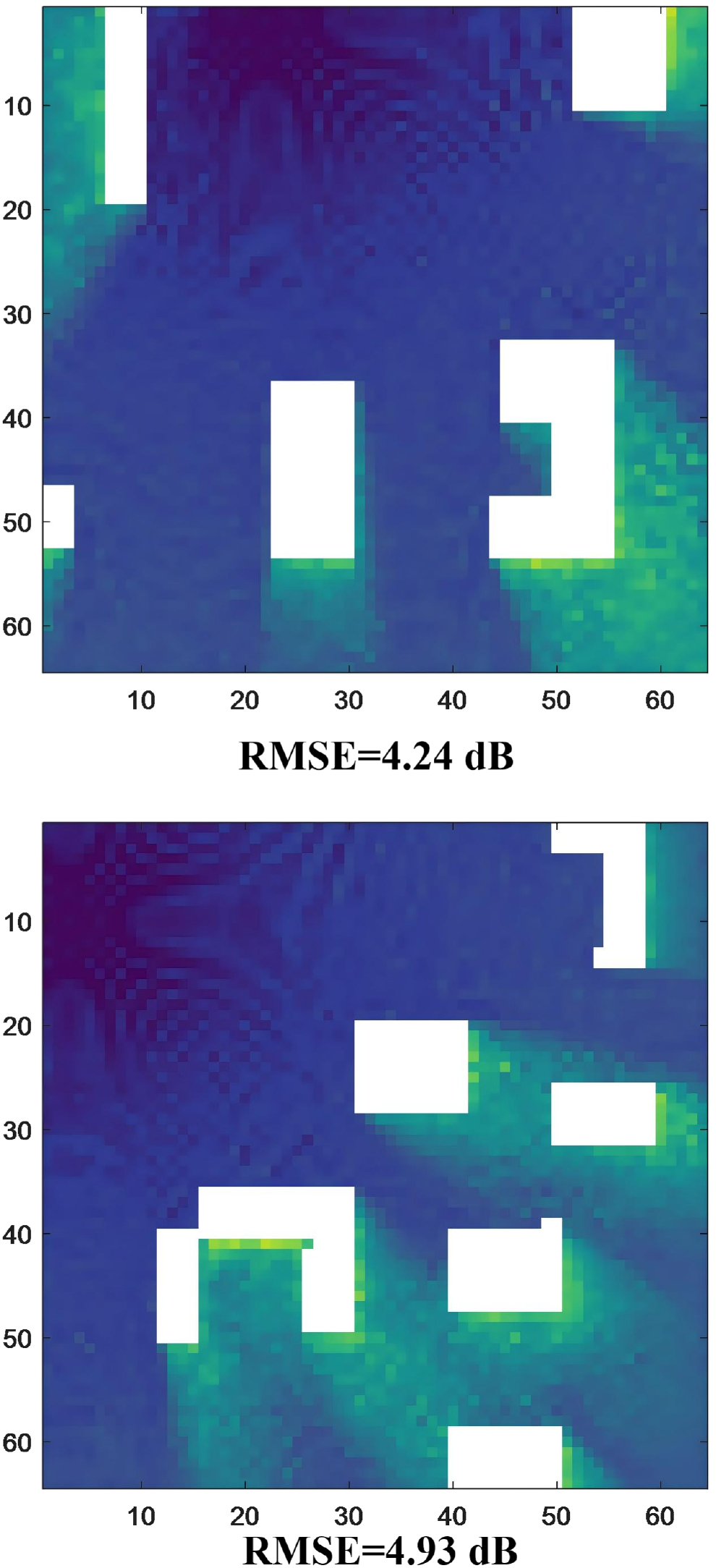}
        \caption{ChannelLM}
        \label{FigPLMapPreFigurec}
    \end{subfigure}
    \hfill
    \begin{subfigure}[t]{0.265\textwidth}
        \centering
        \includegraphics[width=\textwidth]{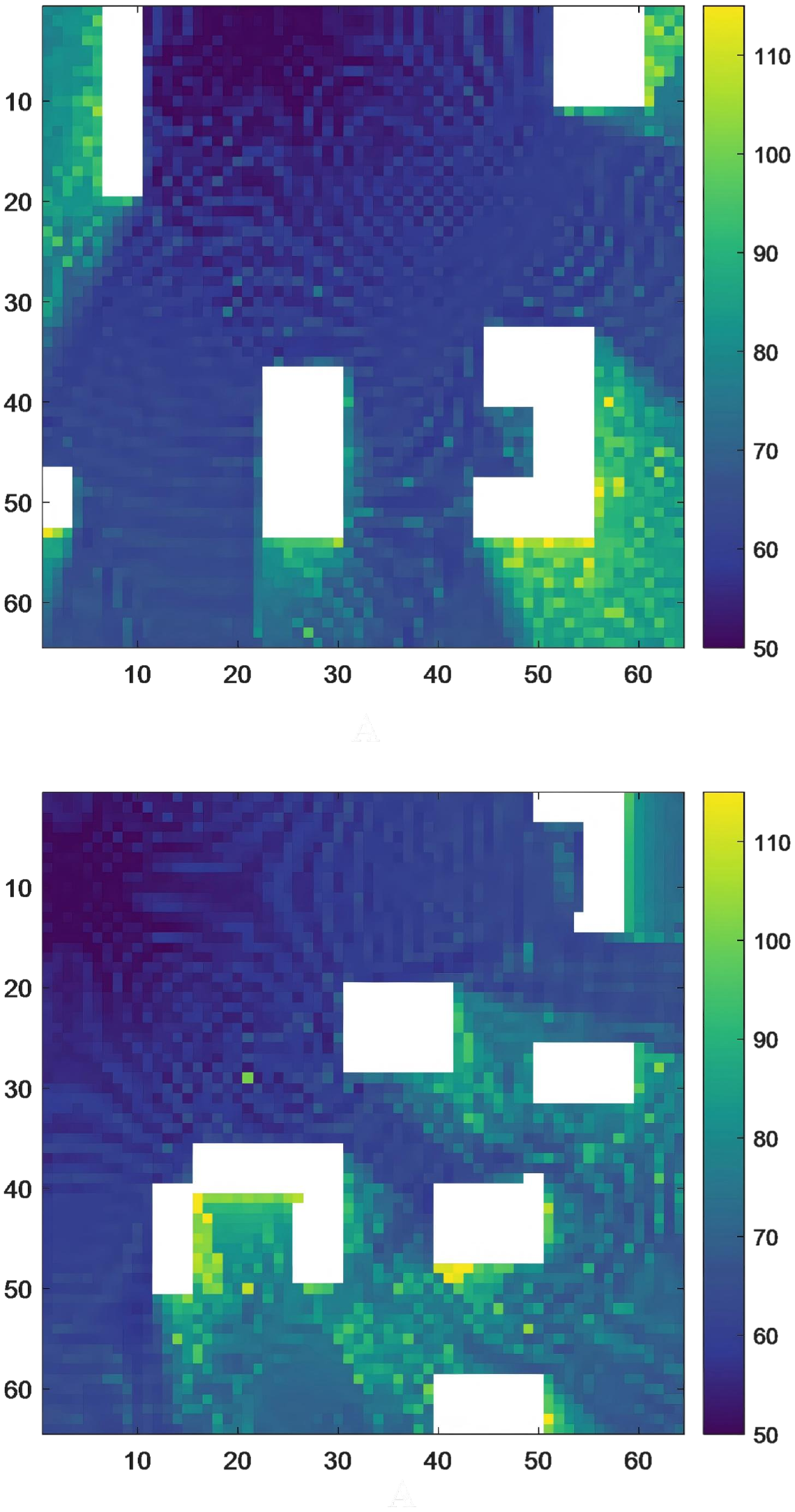}
        \caption{Ground-truth}
        \label{FigPLMapPreFigured}
    \end{subfigure}
    \caption{The visual comparison of PL maps prediction.}
    \label{FigPLMapPreFigure}
\end{figure*}

In addition, Fig. \ref{FigPLMapPreFigure} presents the visual comparison of PL maps prediction on two representative test samples. From the perspective of overall spatial consistency, the prediction results of ChannelLM exhibit the closest global structural resemblance to the ground-truth PL maps. In contrast, both W/o GPT-2 and PMNet show varying degrees of spatial structural distortion. The W/o GPT-2 model, which consists of only three CNN layers, has a limited effective receptive field and lacks the capability for cross-region information propagation. Consequently, it struggles to adequately diffuse global prior information about BS locations across the entire spatial domain and to establish a continuous, coherent propagation structure. As a result, its predictions manifest distinct local anomalies near the BS (e.g., the clearly bounded square region in the upper-left corner of the figure) and fail to accurately reconstruct the increasing trend of PL with distance in LoS regions. 

PMNet adopts a deep U-Net architecture, which integrates multi-scale feature extraction with skip connections to combine high-level semantic information with low-level spatial details. This design expands the effective receptive field and mitigates the limitations associated with relying solely on local convolution operations. Thus, PMNet can better restore the overall trend of PL variation within LoS regions. However, its predictions still deviate from the ground-truth in the distribution of shadow regions caused by building obstructions. In comparison, ChannelLM not only more accurately reconstructs the gradual variation of PL in LoS regions and the detailed texture around the BS, but also effectively recovers the position and shape of shadowed regions behind buildings. Furthermore, it demonstrates smoother and more reasonable transitions at the boundaries between LoS and NLoS areas.

\begin{table}[t]
\centering
\caption{CSI matrix prediction results in training and unseen new test environments.}
\label{TableCSIResult}
\renewcommand{\arraystretch}{2.0}
\begin{tabular}{|c|c|c|c|c|}
\hline
\multirow{2}{*}{\textbf{Model}} & \multicolumn{2}{c|}{\textbf{Training environments}} & \multicolumn{2}{c|}{\textbf{New test environments}} \\
\cline{2-5}
 &\textbf{\makecell{NMSE\\(dB)}} &\textbf{SGCS}&\textbf{\makecell{NMSE\\(dB)}} &\textbf{SGCS} \\
\hline
\makecell[c]{Small model\\ (w/o env. features)} & -8.96 & 0.8757 & -8.91 & 0.8754 \\
\cline{1-5}
ChannelLM              & -13.49 & 0.9580 & -13.13 & 0.9554 \\
\hline
\end{tabular}
\end{table}


In CSI matrix prediction, ChannelLM is compared with its ablated version without environment features (W/o environment feature), and the NMSE and SGCS results on the test set are summarized in Table~\ref{TableCSIResult}. The results show that ChannelLM significantly outperforms the ablated variant in both metrics. Specifically, ChannelLM achieves an NMSE reduction of 4.53 dB in training environments and 4.23 dB in unseen new test environments, demonstrating that incorporating environment features effectively improves the model’s accuracy in recovering both large‑scale and small‑scale information from the CSI matrix. Meanwhile, ChannelLM also attains a better SGCS score, with an improvement of approximately 8\% across both training and unseen new test environments, indicating its stronger capability in predicting the small‑scale phase information of CSI. These results collectively confirm that the environmental modeling enhances the prediction performance of CSI.

\begin{figure}[htbp]
\centerline{\includegraphics[width=0.5\textwidth]{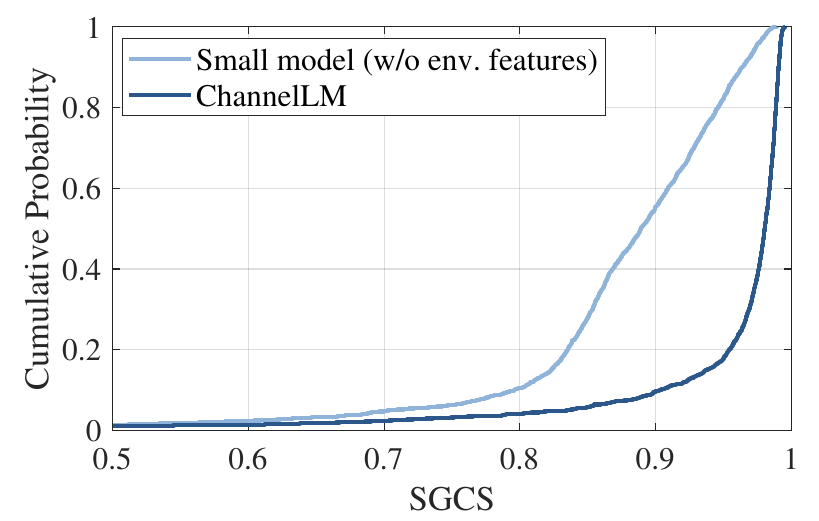}}
\caption{CDF of CSs between the predicted and the ground-truth CSI.}
\label{FigCSICSCDF}
\end{figure}
Fig. \ref{FigCSICSCDF} presents the cumulative distribution function (CDF) of the SGCS for the proposed ChannelLM and small model variant (w/o env. features) in unseen new test environments, offering further insight into the distribution characteristics of the prediction results. At the 80\% CDF point, the corresponding SGCS values are 0.9461 and 0.9886, respectively. Nevertheless, the CDF curve of ChannelLM is consistently shifted to the right compared to that of the ablated version, indicating that for the vast majority of test samples, the predicted CSI exhibits higher structural consistency with the ground-truth CSI.

\begin{figure}[htbp]
\centerline{\includegraphics[width=0.5\textwidth]{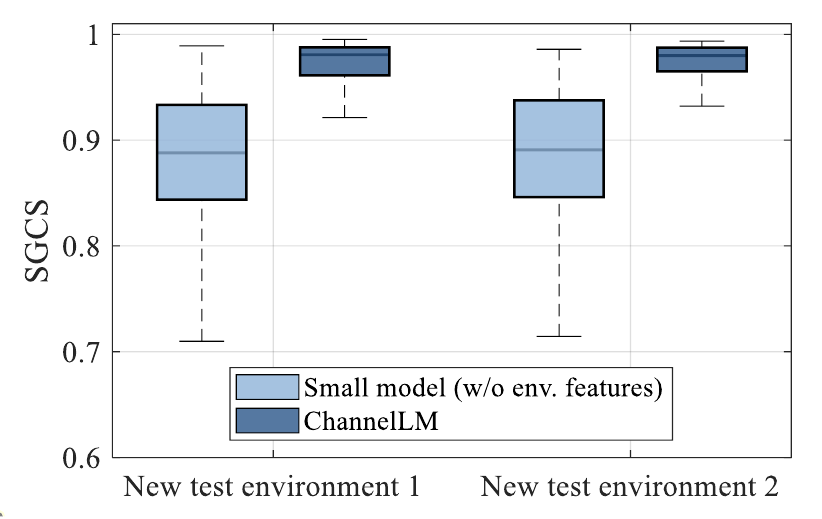}}
\caption{The SGCS distribution across two unseen new test environments.}
\label{FigCSICS}
\end{figure}
Fig. \ref{FigCSICS} further compares the SGCS distributions of the proposed ChannelLM and its ablated version small model (w/o env. features) across two unseen new test environments. The observed diversity in SGCS distribution between the environments indicates variations in their respective prediction difficulty. Nonetheless, ChannelLM consistently demonstrates superior performance in both cases. This confirms that its advantage is not contingent upon a specific test environment but remains stable under diverse spatial structures and propagation conditions.

\vspace{4mm}
\noindent\textbf{System-Level Evaluation in Dynamic Scenarios}

\begin{figure*}[t]
\centerline{\includegraphics[width=1\textwidth]{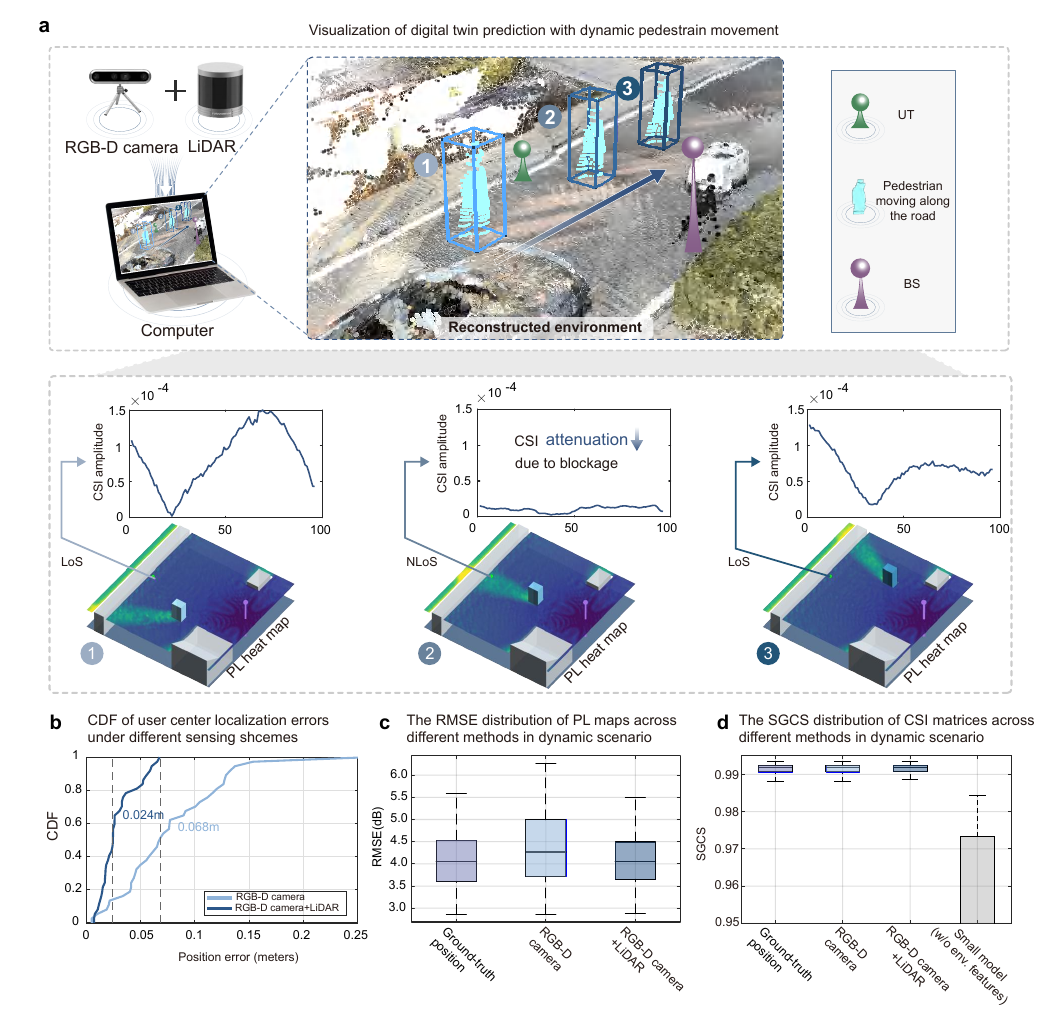}}
\caption{Digital twin prediction results in a dynamic environment induced by pedestrian movement.}
\label{DynResult}
\end{figure*}

In practical applications, the ChannelLM architecture relies on the coordination of multiple functional modules. Therefore, a system-level evaluation is conducted in simulated scenarios with dynamic pedestrians to comprehensively assess the complete pipeline of sensing, feature extraction, and digital twin prediction.


To evaluate sensing and environment reconstruction accuracy in real-world scenarios, the localization error between the estimated user center position obtained by the sensing system and the ground-truth user center position is adopted as the evaluation metric. Under varying distance conditions, 37 test locations are selected. At each location, mobile targets are identified, and their 2D center positions are estimated using two sensing schemes: a single-modal scheme based on an RGB-D camera and a multi-modal scheme that fuses RGB-D camera and LiDAR measurements.

Fig.~\ref{DynResult} (b) shows the CDF of the user center localization error for the two sensing schemes. The single-modal sensing scheme exhibits relatively large localization errors, with some test samples exceeding 10 centimeters (cms). In contrast, the multi-modal sensing scheme consistently reduces the localization error across all evaluated positions.


From a quantitative perspective, the median localization errors of the single-modal and multi-modal sensing schemes are 6.8 cms and 2.4 cms, respectively. Compared with the single-modal sensing approach, the multi-modal sensing scheme achieves an error reduction of approximately 65\%. These results demonstrate that fusing complementary information from heterogeneous sensors significantly improves the robustness and accuracy of user center localization, particularly in scenarios with longer sensing distances. More importantly, the above results validate the reliability of environment sensing in real-world scenarios, which is a critical prerequisite for constructing high-fidelity 3D environment models and extracting environment features for subsequent digital twin prediction in ChannelLM architecture.


Based on the above localization results, we further analyze digital twin prediction performance in dynamic scenarios by examining the impact of localization errors. Specifically, the evaluation is conducted on 97 samples constructed from the 37 tested pedestrian locations. Among them, 37 samples correspond to single-pedestrian scenarios, while the remaining 60 samples correspond to two-pedestrian scenarios formed by randomly selecting two locations from the same set of 37 positions. Their effects on PL map prediction and CSI matrix prediction are evaluated separately over these 97 samples.

PL map prediction relies entirely on environment feature inputs and, therefore, is more sensitive to the accuracy of environment information. Under this setting, localization errors directly affect the spatial placement of dynamic targets in the reconstructed environment, leading to shifts in environment features such as scatterer locations, blockage relationships, and NLoS regions. These discrepancies are ultimately reflected as PL map prediction errors. Fig.~\ref{DynResult} (c) illustrates the RMSE distributions of PL prediction under different localization schemes. Benefiting from more accurate and stable target localization, the multi-modal scheme maintains higher geometric consistency in the reconstructed environment, thereby achieving PL prediction performance close to the ideal case without localization errors and reducing the RMSE by 0.28 dB compared with the single-modal sensing scheme.

In contrast to PL prediction, CSI matrix prediction incorporates not only environment information but also partial pilot CSI as observation constraints, making it relatively less sensitive to localization errors. On the one hand, pilot observations provide direct measurement cues for channel reconstruction, enabling the model to preserve effective recovery of channel amplitude and phase structures even when slight deviations exist in the environmental prior. On the other hand, the environment information used by the CSI branch is not directly fed in raw geometric form but is represented through high-level abstract features extracted and fused by the network, which exhibit a certain degree of robustness to small-scale positional perturbations. As a result, under the current sample setting, the impact of localization errors on CSI prediction performance remains limited, manifesting only as a mild increase in NMSE and a slight degradation in SGCS, as shown in Fig.~\ref{DynResult} (d). Finally, we further compare ChannelLM with its small model variant (w/o env. features). The results show that removing environment features leads to a significant increase in CSI prediction error.
    

Fig.~\ref{DynResult} (a) further presents a visual example of digital twin prediction for a representative sample when a pedestrian moves along the road, where the PL map is shown as a heat map, while the CSI is presented in terms of amplitude. As the pedestrian position changes, the corresponding non-line-of-sight (NLoS) region behind the pedestrian varies accordingly in the predicted PL map. Meanwhile, the CSI exhibits noticeable fluctuations due to the dynamic blockage effect introduced by pedestrian movement. This observation further confirms that the proposed architecture is able to reflect the dynamic influence of pedestrian movement on both large-scale and small-scale channel fading. The quantitative prediction results over all evaluated samples are summarized in Table~\ref{TableDYNResult}. Several observations can be drawn from these results. Localization errors in the sensing stage mainly affect channel prediction by degrading the geometric consistency of environment features. This impact is more direct for PL prediction tasks that rely exclusively on environmental inputs, while it is relatively weaker for CSI prediction tasks that jointly leverage pilot observations and abstract environmental representations. Meanwhile, environmental information itself remains a key factor in improving CSI prediction accuracy. In addition, the prediction errors in this subsection are generally smaller than those in the static-scenario evaluation, likely because the considered dynamic scenes contain fewer scatterers than the static simulation scenarios, resulting in a less complex propagation environment.

\begin{table}[htbp]
\centering
\caption{PL and CSI prediction results in dynamic scenario.}
\label{TableDYNResult}
\renewcommand{\arraystretch}{1.5}
\begin{tabular}{|c|c|c|c|}
\hline
\rule{0pt}{4ex}\textbf{Method} & \textbf{\makecell[c]{RMSE of PL\\ map (dB)}} &\textbf{\makecell[c]{NMSE of \\ CSI matrix (dB)}} &\textbf{\makecell[c]{SGCS of \\ CSI matrix}} \\
\hline
True position & 4.06 & -19.14 & 0.9870  \\
\hline
RGB-D camera& 4.35 & -19.10 & 0.9869 \\
\hline
\rule{0pt}{4ex}{\makecell[c]{RGB-D camera\\ + Lidar}}& 4.07 & -19.14 & 0.9870 \\
\hline
\rule{0pt}{4ex}{\makecell[c]{Small model\\ (w/o env. features)}}& - & -11.70 & 0.9279 \\
\hline
\end{tabular}
\end{table}

In addition, the end-to-end inference latency of the proposed system is measured and analyzed. During the sensing and environment reconstruction stage, YOLO-based dynamic object detection using RGB images takes approximately 45.98 ms, coarse localization based on depth images requires about 0.56 ms, and clustering and fine localization using point clouds by the density-based spatial clustering of applications with noise (DBSCAN) algorithm \cite{dbscan} take around 8.38 ms. The total latency of this stage is approximately 54.93 ms, which constitutes the dominant portion of the overall system latency. In the environment feature extraction stage, the extraction of the height map takes about 0.30 ms, while the computation of the penetration ratio map requires approximately 0.87 ms, resulting in a total latency of 1.17 ms. Subsequently, the ChannelLM-based prediction of the PL map and CSI matrix takes approximately 18.57 ms. By aggregating the latencies of all stages, the overall system-level inference latency in simulation is around 70 ms, indicating that the proposed ChannelLM-driven DTC architecture is capable of supporting real-time inference requirements for DTC online applications.

\vspace{4mm}
\noindent\textbf{Discussion}

This paper proposes a ChannelLM-driven DTC architecture for environment-generalizable channel prediction in the 6G AI-enabled air interface. The proposed architecture consists of three components. First, environment reconstruction is achieved through multimodal sensing with RGB-D cameras and LiDAR. Second, radio propagation-guided environment features are extracted, including physically interpretable BS/UT location maps, scatterer distribution and height maps, and penetration ratio maps. Finally, large-scale and small-scale channel parameters are predicted through ChannelLM.

Simulation results demonstrate that, in previously unseen test environments, the proposed ChannelLM-driven DTC architecture achieves superior PL map prediction accuracy and CSI reconstruction performance compared with conventional U-Net-based models and ablated small models, indicating enhanced cross-scenario generalization capability. System-level evaluations in dynamic scenarios with pedestrian movement further show that multimodal sensing significantly reduces localization errors and improves the geometric consistency of reconstructed environments, thereby benefiting subsequent propagation prediction tasks. Moreover, the overall inference latency of approximately 70 ms confirms the feasibility of real-time deployment. This architecture enables real-time joint prediction of large-scale and small-scale channel fading in dynamic environments, thereby providing effective support for the AI-enabled air interface.

\vspace{4mm}
\noindent\textbf{Methods}

This section provides a comprehensive elaboration on the implementation details of the ChannelLM-driven DTC architecture illustrated in Fig. \ref{FigChannelLMFramework}. It specifically covers multi-modal sensing-based environment reconstruction, radio propagation-guided environment feature extraction, the detailed design of the channelLM core, dataset construction, and the training and validation configurations of ChannelLM.

\vspace{2mm}
\noindent\textbf{Multi-Modal Sensing-Based Environment Reconstruction}
\begin{figure}[t]
\centerline{\includegraphics[width=0.45\textwidth]{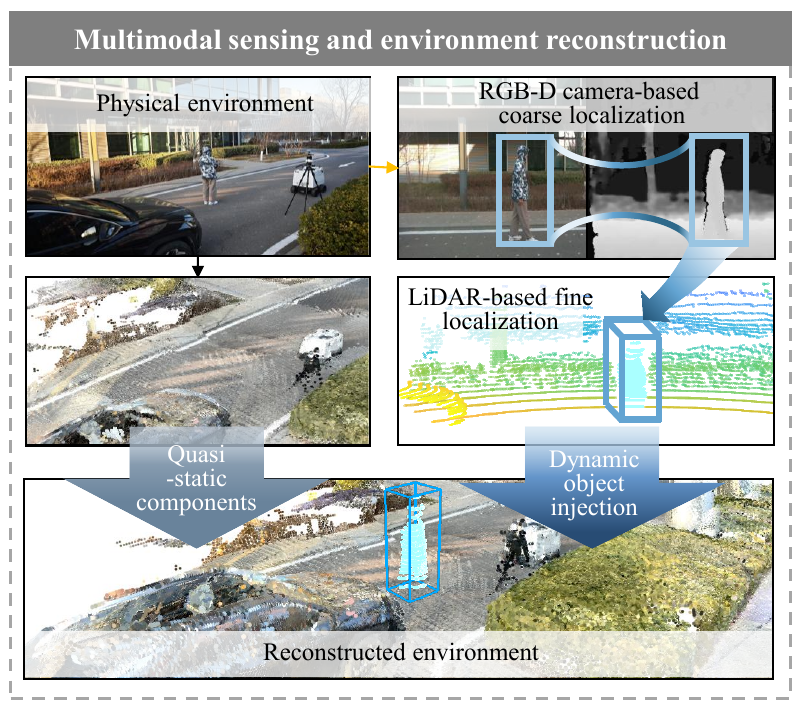}}
\caption{Environment reconstruction framework integrating quasi-static components and online dynamic object injection.}
\label{EnvRecResult}
\end{figure}

The proposed reconstruction procedure consists of two complementary parts, namely quasi-static environment component reconstruction and dynamic object reconstruction, as shown in Fig.~\ref{EnvRecResult}. For quasi-static environment components, including buildings, roads, and vegetation, that are temporally stable and can therefore be modeled in high‑fidelity 3D via a one‑time offline measurement. The resulting model serves as a persistent static reference for online environment reconstruction, avoiding repeated updates and substantially lowering the computational burden in real‑time operation. The reconstruction workflow consists of three main steps: selection of the target communication region; acquisition of the 3D geometry and the relative spatial arrangement of all static scatterers within the region using a laser rangefinder; and high‑accuracy reconstruction of the static environment model with the open‑source 3D modeling software Blender.

For dynamic object reconstruction, this paper employs a multi-modal sensing system comprising an RGB-D camera and a LiDAR to identify and localize dynamic objects. The RGB-D camera can simultaneously capture pixel-aligned RGB images and depth information, outputting dense depth maps in which each pixel contains 3D coordinates and corresponding RGB color values. However, the effective range of this sensor is typically limited (generally from a few meters to over ten meters), and it is susceptible to interference from complex lighting conditions such as strong direct light or backlighting~\cite{rgbd}. 

\begin{figure}[t]
\centerline{\includegraphics[width=0.5\textwidth]{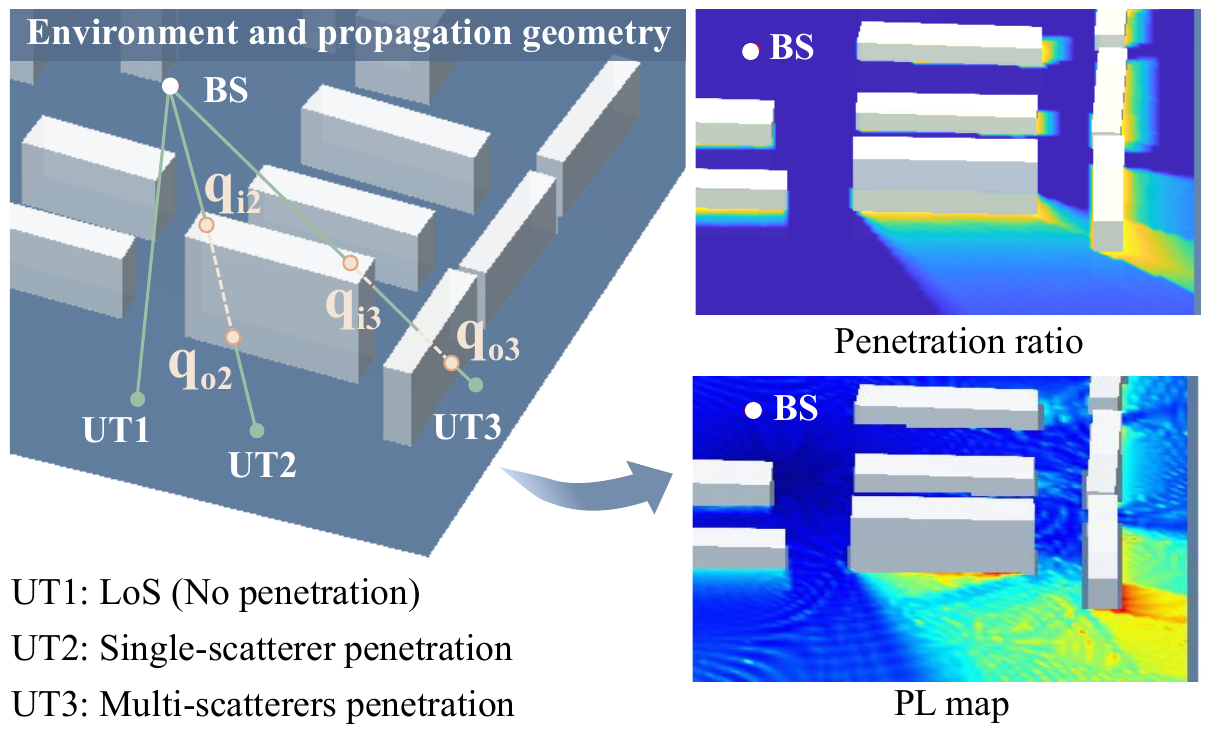}}
\caption{Calculation of penetration ratio for three different cases.}
\label{FigPenRatio}
\end{figure}

In contrast, LiDAR can directly acquire high-precision 3D point cloud data, with each point containing 3D coordinate information. It offers advantages such as long detection range, high accuracy, and strong resistance to lighting interference, yet it cannot directly provide color or texture features of the environment, which limits its ability to interpret semantic information of the environment~\cite{lidar}. To address this limitation,  this paper fuses data from both sensors. Specifically, RGB-D images are first used to identify dynamic objects and obtain their coarse positions. Based on these coarse localization results, LiDAR point clouds in the neighboring region of each target are further selected to perform fine localization. In this way, the RGB-D camera provides efficient target detection and coarse localization, while LiDAR point clouds enable more accurate position refinement. As a result, the sensing accuracy and system robustness in dynamic scenes are improved.

The dynamic environment reconstruction algorithm proposed in this paper consists of four main steps. Firstly, based on the RGB images provided by a depth camera, the advanced YOLOv11 algorithm~\cite{yolo} is employed to identify the categories of scatterers and mark the corresponding 2D pixel regions with rectangular bounding boxes. Subsequently, leveraging the correspondence between RGB pixels and 3D coordinates, the approximate 3D center coordinates and rough orientation of the target within each bounding box are estimated for RGB-D camera-based coarse localization. Next, using the obtained 3D center coordinates as a reference, higher-precision LiDAR point clouds in the vicinity of the target are selected and clustered via the DBSCAN algorithm~\cite{dbscan} for LiDAR-based fine localization, to extract more accurate 3D center positions and boundaries of the dynamic objects. Finally, based on the clustered point clouds, the center points and normal vectors of each surface of the bounding box are calculated, simplifying the dynamic objects into cuboid geometric models. Specifically, for the communication scenario considered in this paper, dynamic scatterers primarily include pedestrians and vehicles. Accordingly, the focus of detection and recognition in the aforementioned dynamic environment reconstruction process is also concentrated on these two types of objects.

\vspace{4mm}

\noindent\textbf{Environment Feature Extraction}

The environment features extracted in this work encompass three categories: BS/UT location maps, scatterer distribution and height maps, and penetration ratio maps. All maps have a spatial resolution of 0.1 meters, meaning the pixel spacing corresponds to a physical distance of 0.1 meters in the actual communication environment. The first two types of features are relatively straightforward and can be directly obtained by equation \eqref{BBSUTq} and \eqref{Hq}, respectively. As illustrated in Fig. \ref{FigPenRatio}, the derivation of the third feature (the penetration ratio) involves three distinct cases.

When no scatterer exists on the BS-UT line, corresponding to the communication situation between BS and UT1 in Fig. \ref{FigPenRatio}. Due to the presence of a LoS component in the wireless signal propagation, the penetration ratio is set to 0 in accordance with the definition given in equation \eqref{PRq}.

When a single scatterer is located on the BS-UT line, corresponding to the communication situation between BS and UT2 in Fig. \ref{FigPenRatio}. Due to a single scatterer located on the line connecting BS and UT2, parameters $\mathbf{q}_i$, $\mathbf{q}_o$ and $\mathbf{q}$ in equation \eqref{PRq} correspond to $\mathbf{q}_{i2}$, $\mathbf{q}_{o2}$ and $\mathbf{q}_{UT2}$ in Fig. \ref{FigPenRatio}, respectively. Specifically, $\mathbf{q}_{UT2}$ denotes the 3D coordinate of UT2. $\mathbf{q}_{i2}$ and $\mathbf{q}_{o2}$ denote the intersection coordinates of the scatterer with the BS-UT2 line, closest to the BS and UT2, respectively.

\begin{figure*}[t]
\centerline{\includegraphics[width=1.0\textwidth]{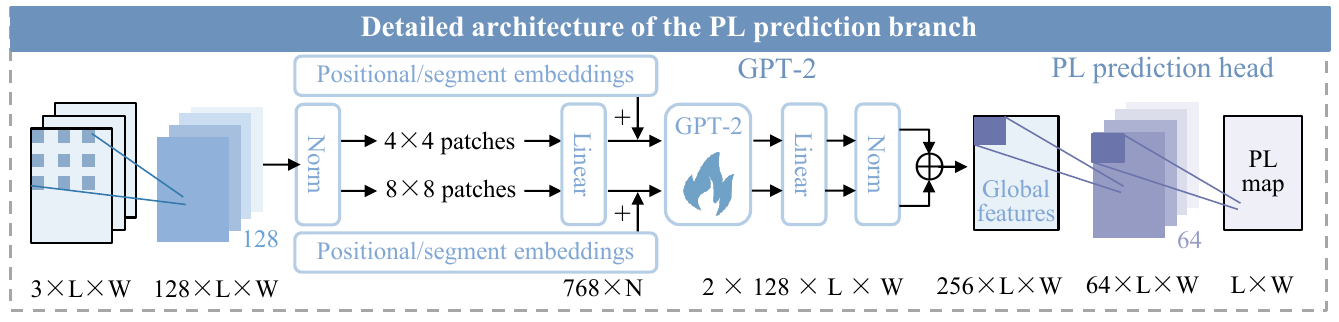}}
\caption{The details of PL map prediction network.}
\label{FigPLPreNet}
\end{figure*}

When multiple scatterers are located on the BS-UT line, corresponding to the communication situation between BS and UT3 in Fig. \ref{FigPenRatio}. Due to multi-scatterers located on the line connecting BS and UT3, parameters $\mathbf{q}_i$, $\mathbf{q}_o$ and $\mathbf{q}$ in equation \eqref{PRq} correspond to $\mathbf{q}_{i3}$, $\mathbf{q}_{o3}$ and $\mathbf{q}_{UT3}$ in Fig. \ref{FigPenRatio}, respectively. Specifically, $\mathbf{q}_{UT3}$ denotes the 3D coordinate of UT3. $\mathbf{q}_{i3}$ and $\mathbf{q}_{o3}$ denote the coordinates of the intersection points between the BS-UT3 line and all scatterers, closest to the BS and UT3, respectively.

Furthermore, the right side of Fig. \ref{FigPenRatio} presents the spatial distributions of the penetration ratio and the PL within the selected communication area. It can be seen that when the UT is close to scatterers, the penetration ratio exhibits considerable fluctuations, thereby exerting a substantial influence on the local PL.

\vspace{4mm}
\noindent\textbf{Details of ChannelLM for PL Map Prediction}

This part presents the detailed network architecture design for digital twin prediction, specifically targeting the PL map and CSI matrix. For PL map prediction, the overall architecture of the proposed network is illustrated in Fig. \ref{FigPLPreNet}. As depicted from left to right, the network consists of the following components: input environment feature maps (including the BS location map, scatterer distribution and height map, and penetration ratio map), a GPT‑2‑based environment feature encoder, a PL prediction head, and the final predicted PL map output. 

Next, the specific design principles of the GPT-2-based environment feature encoder and PL prediction head will be elaborated in detail. The proposed GPT-2-based environment feature encoder first preprocesses three-channel inputs through a dilated convolutional layer, followed by the construction of dual-scale patch branches for multi-scale feature extraction. The features are then globally fused by GPT-2 to generate a 256-channel global environment feature. Specifically, the input feature maps are initially passed through a dilated convolution layer with kernel size=5, padding=4, and dilation=2, projecting them into a 128-dimensional unified feature space to enhance local structural representation. 

To capture both fine-grained geometric details and global spatial structures, two parallel patch branches are constructed: a small-scale branch employs $4\times 4$ patches with an overlap of 1 to retain subtle local variations, while a large-scale branch adopts non-overlapping $8\times 8$ patches to encode broader spatial contexts. Both branches project features to the GPT-2 hidden dimension of 768 via linear layers, incorporating learnable 2D positional encodings and using 0/1 segment embeddings to distinguish between scales. The resulting multi-scale token sequences are fed into GPT-2 for global relational modeling. The outputs of each branch are then linearly mapped back to their corresponding 2D spatial structures, where overlapping regions in the small-scale branch are averaged to suppress boundary artifacts. Finally, the features from both branches are concatenated along the channel dimension, yielding a 256-dimensional global environment feature. 

The design motivation for the preceding CNN structure draws from research on Vision Transformer (ViT)~\cite{VIT}. In pure Transformer architectures, attention heads in lower layers often focus only on local regions, and introducing a CNN feature extraction module helps mitigate such locality biases. This hybrid design provides more stable local structural priors for the Transformer, thereby reducing its burden of learning low-level patterns. The author in~\cite{VIT} indicates that this hybrid architecture achieves better performance and generalization with smaller model sizes, showing significant gains at 86M parameters, while benefits diminish at scales of 307M and 632M parameters. Based on this observation, the proposed method incorporates the aforementioned dilated convolution layer before GPT-2 Small (124M parameters), effectively decoupling local and global features within a constrained parameter budget.

The PL prediction head adopts a dual-layer lightweight convolutional architecture to map the global geometric features output from the GPT-2-based environment feature encoder into the final PL map. The first layer employs a $5\times 5$ convolutional kernel to perform spatial smoothing, while the second layer uses a $3\times 3$ kernel to compress the channels into a single-channel output. Both convolutional layers maintain the spatial resolution unchanged through padding, ensuring that the decoded result is strictly aligned with the input environment features at the pixel level. The dimensional configuration of each hidden layer is illustrated in Fig. \ref{FigPLPreNet}.

\vspace{4mm}
\noindent\textbf{Details of ChannelLM for CSI Prediction}
\begin{figure*}[t]
\centerline{\includegraphics[width=1.0\textwidth]{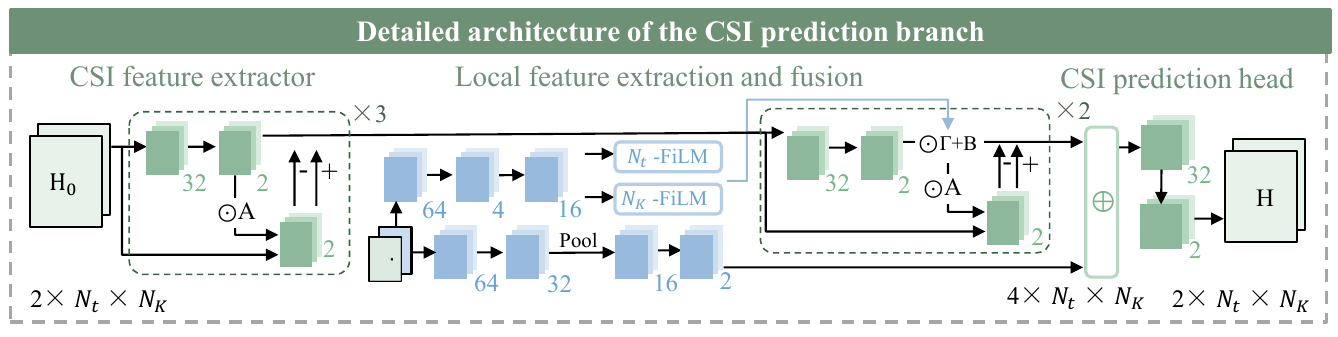}}
\caption{The details of CSI matrix prediction network.}
\label{FigCSIPreNet}
\end{figure*}

For CSI matrix prediction, Fig. \ref{FigCSIPreNet} illustrates the overall architecture of the proposed CSI matrix prediction network. As depicted from left to right, the network comprises the following modules: the lightweight pilot channel observations as input (with real and imaginary parts separated to form a dual-channel image), the CSI feature extractor, the local feature extraction and fusion module (which incorporates the UT position map as an additional input), the CSI prediction head, and finally the predicted CSI matrix as the output. Next, the specific design principles of the CSI feature extractor, local feature extraction and fusion module, and CSI prediction head will be elaborated in detail.

The CSI feature extractor, inspired by~\cite{DPS}, employs a CNN to approximate proximal gradient iterations~\cite{PROX}, mapping partially estimated CSI from lightweight pilot signals into CSI latent features. Specifically, the iterative procedure is successively implemented by CNNs, with the update at the $k-$th iteration can be expressed as follows:
\begin{equation}   
    \mathbf{Z}_{k}=\mathbf{H}_{LF,k-1}-\beta_{\theta,k}\left(\mathbf{A}\odot\mathbf{H}_{LF,k-1}-\mathbf{H}_{0}\right),
    \label{CNNAPG}
\end{equation}    
\begin{equation}       \mathbf{H}_{LF,k}=\mathcal{P}_{\theta,k}\left(\mathbf{Z}_{k}\right),
    \label{CNNAPG}
\end{equation}
where $\mathbf{H}_{LF,k-1}$ and $\mathbf{H}_{LF,k}$ are the $k-1$-th and $k$-th CSI latent features, respectively. $\mathbf{H}_{0}$ is the partially estimated CSI from lightweight pilot signals. $\mathbf{Z}_{k}$ is the intermediate proximal gradient update. $\mathcal{P}_{\theta,k}$ and $\beta_{\theta,k}$ are the proximal mapping and the stepsize at the $k$-th iteration, which are modeled using a two-layer CNN and a single-layer CNN, respectively. $\mathbf{A}$ is the sampling matrix defined in equation \eqref{CSIMatrixZero}. Besides, the number of iteration is set to 3 in this paper.

The local feature extraction and fusion module is designed to characterize the local propagation properties around a specific UT and enhance the reconstruction accuracy of the CSI matrix. It integrates the global scene structure with the UT location map to extract local environment features, and further employs a feature-level fusion mechanism to combine these local features with the CSI latent features produced by the CSI feature extractor. Specifically, in the local feature extraction stage, the two-scale global environment features previously extracted by the GPT-2 backbone are first aggregated via element-wise summation, and then concatenated with the UT location map along the channel dimension, resulting in a 129-dimensional input. A CNN network is subsequently employed to extract local environment features corresponding to the UT’s neighboring region. 

In the feature fusion stage, feature-wise linear modulation (FiLM)~\cite{film} is introduced into the proximal gradient unrolling process as the fusion mechanism. The local environment features are used to generate a set of modulation parameters, which are applied to the CSI latent features $\mathbf{H}_{LF,k}$ through element-wise scaling and shifting operations. In this manner, the environmental priors are explicitly injected into the channel reconstruction process without disrupting the original proximal unrolling structure. Specifically, the local environment features are aggregated along the two spatial directions, and linear layers are utilized to generate modulation parameters for the antenna dimension $N_t$ and the subcarrier dimension $N_K$, respectively. These parameters are then combined via an outer-product operation to form FiLM coefficients that cover the two-dimensional antenna–subcarrier grid. The conditional modulation of the intermediate variable $\mathbf{H}_{LF,k}$ can thus be expressed as:
\begin{equation}
\tilde{\mathbf{H}}_{LF,k}
= \boldsymbol{\Gamma}_{k} \odot \mathbf{H}_{LF,k}
+ \mathbf{B}_{k},
\end{equation}
where $\boldsymbol{\Gamma}_{k}\in \mathbb{C}^{N_t \times N_{sc}}$ and $\mathbf{B}_{k} \in \mathbb{C}^{N_t \times N_{sc}}$ denote the modulation parameters that control the relative importance of features at different antenna–subcarrier positions and the environment-dependent bias correction, respectively.

In addition, another set of CNNs is employed to extract auxiliary features from the environment information, and a linear interpolation operation is applied to map their spatial resolution to match that of the CSI features, thereby enabling feature concatenation in the spatial domain. This design allows the CSI prediction process to jointly exploit geometric environment priors and the initial channel reconstruction, leading to notable improvements in prediction accuracy and robustness. All convolutional layers in this module preserve the spatial dimensions of the feature maps. Specifically, the environment-related convolutional layers adopt $5\times5$ kernels with padding=2, while the convolutional layers involved in the proximal gradient iterations use $3\times3$ kernels with padding=1. Besides, the number of approximate proximal gradient iterations is set to 2.

The CSI prediction head adopts a lightweight CNN architecture to map the fused features into the final CSI matrix. Specifically, the FiLM-conditioned CSI features $\tilde{\mathbf{H}}_{LF,k}$ and the auxiliary environment features produced by the local feature extraction and fusion module are concatenated along the channel dimension and then processed by a two-layer CNN to generate the CSI prediction. All convolutional layers in this module adopt a $3\times3$ kernel with padding $=1$ to preserve the spatial dimensions of the CSI during decoding.

\vspace{4mm}
\noindent\textbf{Dataset Construction}

To support the training and performance evaluation of ChannelLM, this study constructs 10 outdoor communication scenarios, each with a size of $12.4\times 12.4$ square meters, using the open-source software Blender. The position, number, and height of scatterers in each scenario are randomly assigned to ensure sufficient diversity in the wireless propagation environment of the dataset. These scenarios are then imported into the commercial ray tracing (RT) software Wireless InSite 4.0 developed by REMCOM~\cite{WI}. Through RT simulation, the corresponding multipath channel components for each scenario are obtained. Subsequent processing of these multipath components yields the MISO-OFDM CSI matrices and PL maps. The detailed parameters for channel generation based on Wireless InSite are summarized in Table \ref{TableRT}.
\begin{table}
\centering
\caption{RT simulation parameters in Wireless Insite.}
\label{TableRT}
\begin{tblr}{
  colspec = {c c},  
  hlines,
  vlines,
  row{1-Z} = {m},  
  cell{2-Z}{1} = {c},  
  cell{2-Z}{2} = {c},  
}
\textbf{Parameters}             & \textbf{Value}                                   \\
Carrier frequency               & 7.5 GHz                                          \\
Subcarrier spacing                  & 600 kHz                                            \\
Number of subcarriers                  & 96                                            \\
Number of transmitter antenna                   & 64                                            \\
Number of receiver antenna                   & 1                                            \\
Transmit power                  & 0 dBm                                            \\
Receive power threshold         & -250 dBm                                         \\
Transmit antenna characteristic & Isotropic, vertical polarization                 \\
Receive antenna characteristic  & Isotropic, vertical polarization                 \\
Transmitter location            & Centered, height: 2 m                            \\
Receiver location               & {Uniform coverage,\\height: 1 m, spacing: 0.1 m} \\
RT simulation propagation model & X3D                                              \\
Reflection order                & 6                                                \\
Diffraction order               & 1                                                
\end{tblr}
\end{table}

For any constructed outdoor simulation scenario, the corresponding BS location map, scatterer distribution and height map, and penetration ratio map within the $12.4\times 12.4$ square meters area can be directly extracted based on equations \eqref{BBSUTq}–\eqref{PRq}. Additionally, for each scenario, a complete PL map can be generated from the PL values obtained at all receiver locations in Wireless InSite. To expand the dataset, this paper employs classical data augmentation techniques, including cropping and rotation. The procedure is detailed as follows: First, the original $124\times 124$ pixels image of an arbitrary scenario is cropped with a stride of 4 pixels, producing 256 sub-images of size $64\times 64$ pixels. Subsequently, all sub-images are rotated by $90^\circ$, $180^\circ$, and $270^\circ$, respectively. Through this process, 1024 images are obtained for each communication scenario, comprising BS location maps, scatterer distribution and height maps, penetration ratio maps, and PL maps. 

Furthermore, a UT location is randomly selected within the coverage area of each augmented sample, and the corresponding UT location map, along with the MISO‑OFDM channel CSI matrix, is saved. In total, 10240 samples are collected from 10 communication scenarios. Each sample includes one BS location map, one UT location map, one scatterer distribution and height map, one penetration ratio map, one PL map, and one MISO‑OFDM CSI matrix. In the proposed architecture, the BS location map, scatterer distribution and height map, and penetration ratio map serve as the input to the ChannelLM PL prediction network, with the PL map as its label. The UT location map and a subset of information extracted from specific positions of the MISO‑OFDM CSI matrix (treated as lightweight pilot-estimated CSI) form the input to the ChannelLM CSI prediction network, while the complete MISO‑OFDM CSI matrix acts as the label for CSI prediction.

\vspace{4mm}
\noindent\textbf{Core Configurations for ChannelLM Training and Validation}

\begin{table}
\centering
\caption{Universal training hyper-parameters.}
\label{TableTrain}
\begin{tblr}{
  colspec = {c c},  
  hlines,
  vlines,
  row{1-Z} = {m},  
  cell{2-Z}{1} = {c},  
  cell{2-Z}{2} = {c},  
}
\textbf{Parameters}             & \textbf{Value}                                   \\
GPT2 learning rate (LR)               & $8\times 10^{-4}$                                          \\
PL prediction network LR (other parts) & $8\times 10^{-3}$                                            \\
CSI prediction network LR          & $8\times 10^{-4}$                                            \\
LR gamma, step size         & $0.95$, $10$                                            \\
Batch size                  & $8$                                            \\
Optimizer                  & Adam                                            \\
Training epochs         & $80$                             
\end{tblr}
\end{table}

The proposed ChannelLM is implemented on a single NVIDIA RTX 4090 GPU. Other universal training hyper-parameters are summarized in Table \ref{TableTrain}. The loss function for the PL prediction task is defined as follows. Here, the PL map prediction problem is treated as a 2D image reconstruction task, and therefore a hybrid loss $\mathcal{L}_{PL}$ combining spatial domain supervision and frequency domain supervision~\cite{FFTloss} is adopted.
\begin{equation}
\begin{split}
    {{\cal L}_{PL}} =& \underbrace {\frac{1}{N_{train}}\sum\limits_{n = 1}^{N_{train}} {\rho ({{{\bf{\widehat P}}}_n} - {{\bf{P}}_n})} }_{{\text{Spatial domain loss}}} +\\
    & 0.1 \times \underbrace {\frac{1}{N_{train}}\sum\limits_{n = 1}^{N_{train}} {\rho \left( {\left\| {{\cal F}({{\widehat {\bf{P}}}_n})} \right\| _F- \left\| {{\cal F}({{\bf{P}}_n})} \right\|} _F\right)} }_{{\text{Frequency domain loss}}},
    \label{PLLossFunction}    
\end{split}
\end{equation}
where
$N_{train}$ is the number of PL map in the training sets, $\bf{\widehat P}$ and $\bf{P}$ are the predicted and ground-truth PL map with $64\times 64$ pixels, respectively. $\rho\left(x\right)=\sqrt{x^2+\epsilon^2}$ is the Charbonnier penalty function~\cite{Charbonnier}, where the constant $\epsilon$ is set to 0.001 in this work. $\mathcal{F}\left(\cdot\right)$ is the 2D Fourier transform function.

The spatial domain supervision term $\mathcal{L}_{PL}$ constrains the overall numerical error of the predicted PL map and enables the model to accurately capture the overall fading trend. The frequency domain supervision term is introduced from the perspective of image reconstruction to provide additional supervision on frequency components, especially for preserving local details and mitigating blurring in the predicted map. Therefore, the frequency-domain term is used as an optimization strategy to improve the structural fidelity reconstruction of PL map. Furthermore, the introduced Charbonnier penalty provides smooth and stable gradients as errors approach zero, which improves the stability and robustness of the training process.

The CSI prediction branch is trained using a loss function defined as the MSE between the predicted CSI and the corresponding ground-truth CSI, which is given by
\begin{equation}
\mathcal{L}_{\mathrm{CSI}} = \frac{1}{N_{\mathrm{train}}}
\sum_{n=1}^{N_{\mathrm{train}}}
\left\| \widehat{\mathbf{H}}_{n} - \mathbf{H}_{n} \right\|_F^2,
\end{equation}
where $N_{\mathrm{train}}$ denotes the number of CSI samples in the training set. Meanwhile, $\widehat{\mathbf{H}}_{n}$ and $\mathbf{H}_{n}$ represent the predicted and ground-truth CSI matrices of the $n$-th sample, respectively.

\vspace{4mm}
\noindent\textbf{Data Availability}

The data that supports the findings of this study is available from the corresponding authors upon reasonable request.

\vspace{4mm}
\noindent\textbf{Code Availability}

The codes that support the findings of this study are available from the corresponding authors upon reasonable request.

\vspace{4mm}
\noindent\textbf{Acknowledgements}

This work was supported by the National Natural Science Foundation of China (No. 62525101 and No. 62401084), the National Key Research and Development Program of China (2023YFB2904805), and the Beijing University of Posts and Telecommunications-China Mobile Communications Group Co., Ltd. Joint Institute.

\vspace{4mm}
\noindent\textbf{Author contributions}

Y. C. and Y. Q. contributed equally to this work, jointly designing the algorithm and performing the simulation validation. Y. C., Y. Q., J. Z., L. Y., and G. L. conceived the research idea. Y. C. and Y. Q. wrote the manuscript. J. Z., L. Y., Y. Z., Z. Z., and G. L. revised the manuscript. All authors have read and approved the final version of the manuscript.

\printbibliography

\end{document}